PGP in Data Science

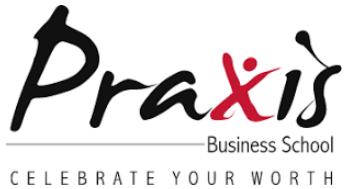

# Precise Stock Price Prediction for Robust Portfolio Design from Selected Sectors of the Indian Stock Market

Capstone project report submitted in partial fulfillment of the requirements for the Post Graduate Program in Data Science at Praxis Business School

By

**Ashwin Kumar R S (C21002)**
**Geetha Joseph (C21003)**
**Kaushik Muthukrishnan (C21004)**
**Koushik Tulasi (C21006)**
**Praveen Varukolu (C21021)**

Under the supervision of

**Prof. Jaydip Sen**
**Professor, Praxis Business School**

# Abstract

Stock price prediction is a challenging task and a lot of research continues to happen in the area. Portfolio construction is a process of choosing a group of stocks and investing in them optimally to maximize the return by minimizing the risk. Beginning from the Markowitz 'Modern Portfolio Theory' a lot of advancement has happened in the area of building efficient portfolios. An investor can get the best benefit out of the stock market if he/she invests in an efficient portfolio and could take the buy/sell decision in advance, by knowing the future asset value of the portfolio with a high level of precision. In this project, we have attempted to build an efficient portfolio and to predict the future asset value by means of individual stock price prediction of the stocks in the portfolio. As part of the project, our team has conducted a study of performance of various statistical, econometric, machine learning and deep learning models in stock price prediction on selected stocks from the chosen five critical sectors of the economy. We have ensured that the validation method used is appropriate for the time series data and have also made some interesting observations regarding the day wise variance of stock price in a week. As part of building an efficient portfolio we have studied multiple portfolio optimization methods beginning from MPT (Modern Portfolio theory). We have built minimum variance portfolio and optimal risk portfolio for all the five chosen sectors by using past five years' daily stock price as training data and have also conducted back testing (next 8 months' data) to check the performance of the portfolio. A comparative study of minimum variance portfolio and optimal risk portfolio with equal weight portfolio is done by backtesting the portfolio. We look forward to continue our study in the area of stock price prediction and portfolio optimization and consider this project as the first step in this regard.

# Table of Contents

# List of Figures

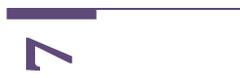

# List of Tables

1 **Chapter 1** | 12

## 1.1 Introduction

The stock market is considered as one of the most lucrative investment options because of its potential to provide huge returns in a short span of time. But at the same time, the stochastic nature of the stock market can cause an investor to suffer a huge loss if he/she isn't adept enough to analyze the market movements. Building an efficient portfolio is one of the ways to protect the investor from suffering huge losses and at the same time ensure profit with some certainty by way of balancing risk and return. Portfolio building is an area that requires thorough financial knowledge and has been the prerogative of asset managers of mutual fund companies for a very long time. But in parallel, studies have been progressing in the statistical, econometric, and data science fields to predict future stock prices and to build optimal portfolios. The pandemic saw a surge in the number of youngsters investing in the Indian stock market (SEBI) and data is considered as the new oil of the 21$^{st}$ century. Enormous data of stock trading is available on the stock exchange websites (NSE and BSE in Indian scenario). In this project, with the help of this freely available data our team has attempted to predict the stock price of selected five stocks from chosen five sectors and to build efficient portfolios for each of the five sectors as an extension to some of the work already done in the area.

A gamut of research has been happening in the area of future stock price prediction. There are two schools of thought related to the feasibility of stock price prediction. The advocates of 'Efficient market Portfolio' argue that stock price prediction is impossible because of the very stochastic nature of the same. But the second group believes that if modeled properly stock prices can be predicted with a good level of accuracy using statistical, econometric, machine learning, and deep learning models. There is a spectrum of research papers published supporting the second school of thought. 'Stock price prediction using Convolutional neural networks on a multivariate time series' (Mehtab & Sen), 'A Robust Predictive Model for Stock Price Forecasting' (Sen & Datta Chaudhuri), 'Stock price prediction using machine learning and deep learning frameworks'(Sen), 'Decomposition of time series data of stock markets and its implications for prediction -An application for the Indian Auto sector' (Sen & Datta Chaudhuri are some of them in this regard.

Building an efficient portfolio is the process of allocating weights to a collection of stocks in such a way that the risk and return are optimized. Markowitz's Minimum Variance Portfolio is considered as the foundation of all the later works in the field of portfolio optimization. Quite a few research papers have been published in the area of portfolio optimization using deep learning models. 'Comparative Analysis of Portfolio Optimization approaches using deep learning models' (Mehtab and Sen), 'Portfolio Optimization on NIFTY Thematic Sector Stocks Using an LSTM Model' (Mehtab, Sen, and Mondal) are a few to list.

In the present work, minimum variance portfolio and optimal risk portfolio are built on five critical sectors of the economy. In each of the five sectors, the top five stocks contributing to the sectoral Index are selected for portfolio building. A detailed comparative study of various statistical, econometric, machine learning, and deep learning models has been done for stock price prediction. Machine Learning models have also been employed for building classification models to predict whether the return for a particular day is positive or negative. Building an optimal portfolio along with the ROI computation of the future asset value using stock price prediction helps an investor to make investment decisions wisely.

The investor can optimally divide the corpus of the fund into a collection of stocks and can take the buy/ sell decision at the right moment to ensure high returns.

The rest of the report is organized as follows. In Chapter 2, the methodology followed in our work is explained. Chapter 3 discusses the statistical and econometric models used for stock price prediction. Chapter 4 discusses the machine learning models used for classification and regression. Chapter 5 presents a detailed discussion on the performance of deep learning models. Chapter 6 is about the various portfolios built and their performances. Finally, Chapter 7 concludes the report.

2 **Chapter 2**

## 2.1 Methodology

The first step towards achieving the goal of stock price prediction and portfolio building was to decide on the 5 sectors. Two criteria were taken into consideration for the same – i) the sectors which are critical to the economy ii) the sectors which flourished well during the pandemic. Based on the above-mentioned criteria the sectors chosen are Metal, Pharma, IT, Bank, and Auto. Once the sectors were chosen the next step was to choose the 5 stocks from each sector. For the same, the latest monthly published sectoral Index report was referred to and the top 5 contributors to the index of each sector were chosen. For the comparative study of stock price prediction, two stocks from each sector were chosen whereas for portfolio optimization five stocks from each sector were chosen. The list of the five stocks chosen from each of the five sectors are mentioned below

*Table 2.1:Five stocks each in five sectors*

| S.No | Metal | Pharma | IT | Banking | Auto |
|------|-------|--------|------|---------|------|
| 1 | Tata Steel | Sun Pharma | Infosys | HDFC Bank | Maruti Suzuki |
| 2 | Hindalco | Divi's Lab | TCS | ICICI Bank | Tata Motors |
| 3 | JSW Steel | Dr. Reddy's Laboratories | Tech Mahindra | State Bank of India | Mahindra & Mahindra |
| 4 | Vedanta | Cipla | Wipro | Kotak Mahindra Bank | Bajaj |
| 5 | Adani Enterprises | Lupin | HCL Technologies | Axis Bank | Eicher Motors |

The daily data from 2016 Jan 1 to Aug 27 2021 was fetched using Yahoo Finance API. The five years and 8 months' data are used for testing and training of statistical, econometric, and machine learning models using the walk-forward validation method. For the deep learning models data from Jan 1 2016 to Dec 31 2020 is used for training and the next 6 months' data is used for testing and for portfolio building data from Jan 1, 2016 to Dec 31, 2020 is used for training and the next 8 months' data is used for backtesting. Usually, stock markets work 5 days a week and will be off on Saturday and Sunday. But in the dataset, some of the weekday trading data was missing due to holidays. The missing days were identified and imputed using forward fill. After the imputation, there are 1476 data points.

The variables present in the data imported are (i) date, (ii) open value of the stock, (iv) high value of the stock, (v) low value of the stock, (vi) close value of the stock, and (vii) volume of the stock. 4 more variables are derived from the above variables and they are i) Day of the week ii) Day of the month iii) Month and iv) Range.  Along with the above-mentioned variables NIFTY index is fetched using Yahoo Finance API and used as one of the variables in order to capture the daily market sentiment. The combined information of historical stock prices and market sentiment help to give a more accurate stock price prediction. Thus, there are 9 predictor variables. Based on the type of model there will be variation in the number and type of predictor and target variables that would be mentioned in the respective sessions. The explanation of each of the variables is given below:

I.     Open: Stock price at the opening time of the stock market

II.    High: The highest price point reached during the trading duration on a particular day

III.    Low: The lowest price point reached during the trading duration on a particular day

IV.    Volume: The no of stocks traded on a particular day

V.    Day of the week: 0-4, represents days from Monday to Friday in order

VI.    Day of the month: 1-31, represents the 31 days of a month

VII.    Month: 1-12, represents the 12 months of a year

VIII.    Range: Close price subtracted from Open

IX.    NIFTY50: NIFTY Index

X.    Close: Stock price at the closing time of the stock market

The models used for prediction of close price are the following:

**Statistical model**: Multivariate Regression, MARS (multivariate adaptive regression splines)

**Econometric Models**: ARIMA (Autoregressive Integrated Moving Average), VAR (Vector Autoregression)

**Machine Learning models**: K Nearest Neighbor, Decision Tree, XGBoost, Random Forest, SVM (Support Vector Machine)

**Deep Learning models**: LSTM (Long Short-Term Memory), CNN (Convolutional Neural Network)

Python libraries are used for building statistical, econometric and machine learning models and Keras for building deep learning models.

The validation method used for statistical, econometric and machine learning models is walk-forward validation. There are two variants of walk-forward validation, expanding window and sliding window walk-forward validation. Certain window size for training and test size is decided. The test size chosen is 14 and the train size chosen is 245. In the expanding window validation, with every iteration both the training set and test set moves forward by a fixed number of data points, the size of the training set would keep on increasing whereas the size of the test set would remain constant. Similarly, in sliding window validation with every iteration both training set and test set moves forward, but unlike expanding window the size of the training set and test set remains constant. That is, as the training window moves forward, it would leave the past values.

The behavior of stock prices is related to the recent past value hence the traditional method of train test split wouldn't be appropriate for stock price prediction. The walk-forward validation method allows us to train the model with the recent values. For stock price validation, the sliding window method is considered to be more appropriate compared to the expanding window method as it leaves the past values as the window moves forward. For stock price prediction more than the amount of the data with which a model has trained the recency of the data is important.

3 **Chapter 3**

## 3.1 Statistical and Econometric models

### 3.1.1 Multivariate Regression

Multivariate regression is an extension of multiple regression where there is one dependent variable and more than one independent variable. The model establishes the relationship between independent and dependent variables using a straight line.

Nine variables are used as predictor variables and 'Close price' is the target variable as mentioned in chapter 2. Since 'Day of the week', 'Day of the Month' and 'Month' are categorical variables, they are dummified using the get_dummies() function which led to the addition of 45 more features. Multicollinearity between the variables has been checked before conducting the regression and variables which are collinear are removed. With the remaining variables backward stepwise regression is conducted. Backward stepwise regression is the process by which the regression is started with all the variables and in every step variable which has the least AIC value is removed and the regression is run again. The process is continued until there are no variables to be removed. With the remaining variables, the model is built.

Multivariate regression can predict the close price of a particular day only if the predictor variables of that particular day are available. This often raises questions regarding the practical use of the model in predicting the future value of stock prices. To demonstrate the practical use of Linear Regression, the future values of the predictors are forecasted using ARIMA, and using the forecasted predictor variables the 'Close price 'is predicted.

### 3.1.2   MARS (Multivariate Adaptive Regression Splines)

The approach entails identifying a set of basic linear functions that, when combined, produce the highest prediction performance. MARS is therefore a sort of ensemble of basic linear functions that may perform well on difficult regression problems with numerous input variables and complicated nonlinear interactions. While predicting stock price, one gets to see nonlinear interactions as the time horizon gets smaller and smaller. Therefore, this problem is suitable to be modelled by MARS. The selection of the basis functions is critical to the MARS method. This consists of two stages: the forward-stage, which is the generating phase, and the backward-stage, which is the refining stage.

The forward stage creates candidate basis functions for the model whereas the backward stage removes the model's basis functions. The forward step is to generate basis functions and add to the model. Each value for each input variable in the training dataset is considered a candidate for a basis function, like a decision tree. For the left and right versions of the piecewise linear function of the same split point, functions are always added in pairs. A created pair of functions is only incorporated to the model if it decreases the overall model's error. The backward stage entails removing functions from the model one at a time. A function is eliminated from the model only if it has no effect on performance (neutral) or improves predicted performance.

The change in model performance during the backward step is assessed using cross-validation of the training dataset, often known as generalized cross-validation or GCV. As a result, the influence of each piecewise linear model on the performance of the model may be evaluated. The model's function count is decided automatically, as the pruning process stops when no more improvements can be achieved. The only two important hyperparameters to consider are the total number of candidate functions to produce, which is frequently set to a very large amount, and the degree of the functions to generate. The degree is the amount of input variables that each piecewise linear function considers. This is set to one by default, but it may be increased to allow the model to capture intricate relationships between input variables. The degree is frequently maintained low to keep the model's computing complexity to a minimum (memory and execution time).

The MARS approach has the advantage of only using input variables that improve the model's performance. MARS achieves an automated kind of feature selection, similar to the bagging and random forest ensemble algorithms.

Approach used to predict the stock price using MARS

1.      Close price of the stock is considered from Jan 01,2016 to Aug 27,2021 to build and validate the model

2.      Earth module is imported

3.      max_terms which are the total number of candidate functions to produce during the forward stage is set to 300 and max_degree which is the maximum degree of the functions to generate is set to 3

4.      Using a train data size of 122, 14 days' close price value is forecasted in an iterative manner and the model is validated using sliding and expanding window

### 3.1.3   ARIMA (Autoregressive Integrated Moving Average)

ARIMA is an econometric model used for time series analysis. The AR component of ARIMA indicates that the variable is regressed on its own lagged values whereas the MA part indicates that the regression error is a linear combination of present and past error terms.

ARIMA can be performed only on a stationary series. A series is made stationary by differencing the time series with its lag value. After each differencing, the Augmented Dickey-Fuller (ADF) test is conducted to check the stationarity of the series, and the process is repeated until the series passes the ADF test. The Auto Regression parameter (p), the Difference parameter (d), and the Moving Average parameter (q) are required to fit the ARIMA model to a time series and to perform the univariate forecasting. Python has the auto_arima() function which finds the appropriate p, d, and q value of a series.

'Close price' is the variable used for univariate forecasting. The walk-forward validation method is used for model validation. Using a train data size of 122, 14 days' close price value is forecasted in an iterative manner.

### 3.1.4  VAR (Vector Autoregression)

While predicting stock price, we encounter five major variables which are time series in nature. Those variables are close price, open price, low price, high price, and volume of stocks that are being traded in a particular period. In this case, the open price impacts the closing price, and the link is bidirectional. The aforementioned assertion is valid for every pricing combination. As a result, the Vector Autoregression Model is being investigated to model the pricing. When two or more time-series impact each other, Vector Autoregression (VAR) is a forecasting technique that may be employed. That is, the time series involved have a bidirectional link.

Each variable in the VAR model is described as a linear combination of its own past values and the past values of other variables in the system. Because there are several time series influencing each other, it is treated as a system of equations with one equation for each variable (time series). It is classified as an autoregressive model since each variable (Time

Series) is treated as a function of previous values, implying that the predictors are nothing more than the series' lags (time-delayed values).

Training data: Open, close, high, low price of a stock from Jan 01,2016 to Aug 9,2021

Testing data: Open, close, high, low price of a stock from Aug 10,2021 to Aug 27,2021

**Granger's Causality Test**

It is possible to test whether two or more time-series influence each other, a primary assumption behind VAR through Granger's Causality test. Here open, close, high, low price of a stock from Jan 01, 2016, to Aug 9, 2021, is subjected to a Granger's Causality test. The Granger causality tests the null hypothesis, which states that the coefficients of past values in the regression equation are zero. To put it another way, the previous values of a time series (X) do not influence the other series (Y). So, if the p-value produced from the test is less than the significance level of 0.05, the null hypothesis may be confidently rejected. Ideal Granger's Causation Matrix is an identity matrix.

*Figure 3.1: Granger's Causation Matrix for Divi's Laboratories Stock*

|         | Open_x | High_x | Low_x  | Close_x |
|---------|--------|--------|--------|---------|
| **Open_y**  | 1.0000 | 0.0000 | 0.0000 | 0.0     |
| **High_y**  | 0.0000 | 1.0000 | 0.0000 | 0.0     |
| **Low_y**   | 0.0000 | 0.0000 | 1.0000 | 0.0     |
| **Close_y** | 0.0167 | 0.0015 | 0.0043 | 1.0     |

From the above matrix, it can be inferred that the open, high, low, close price of a stock on a particular day influences each other.

**Co-integration Test**

The co-integration test is used to determine whether or not there is a statistically significant relationship between two or more time-series. The number of differencing necessary to make a non-stationary time series stationary is denoted by order of integration(d). When there is a linear combination of two or more time-series with an order of integration (d) less than that of the individual series, the collection of series is said to be co-integrated. When two or more time-series are co-integrated, it indicates they have a statistically significant relationship in the long run. This is the fundamental principle upon which the VAR model is based.

**Stationarity of Time Series**

Because the VAR model needs the time series you wish to forecast to be stationary, it is common to assess the stationarity of every time series in the system. A stationary time series is one in which the mean and variance do not vary over time. If a series is discovered to be non-stationary, it is made stationary by differencing the series once and repeating the test until it becomes stationary. Because differencing decreases the length of the series by one, and because all the time series must have the same length, one should difference all of the series in the system if one wishes to differ at all. We shall utilize the ADF test to determine stationarity.

*Figure 3.2: ADF Test on Close Price of Divi's Laboratories after performing first-order differencing*

```
        Augmented Dickey-Fuller Test on "Close"
        -----------------------------------------------
Null Hypothesis: Data has unit root. Non-Stationary.
Significance Level    = 0.05
Test Statistic        = -40.4588
No. Lags Chosen       = 0
Critical value 1%     = -3.435
Critical value 5%     = -2.864
Critical value 10%    = -2.568
=> P-Value = 0.0. Rejecting Null Hypothesis.
=> Series is Stationary.
```

**Selecting the order(p) of VAR model**

Order of a VAR model is the number of lags taken into consideration. One of the most significant components of VAR model definition is lag selection. In practice, we often set a maximum number of delays, $p_{max}$, and test the model's performance with $p = 0, 1, 2, \ldots\ldots$

,$p_{max}$. The model VAR(p) that minimizes certain lag selection criterion is thus the optimum model. The following are the most widely used lag selection criteria

- Akaike Information Criterion

- Bayesian Information Criterion

- Hanna Quinn Information Criterion

- Final Prediction Error

During our modeling of stock prices, the number of lags which gave the minimum AIC was selected as the order of the VAR model which will be fitted on the training data.

**Checking for autocorrelation of residuals**

The serial correlation of residuals is used to determine whether or not there is a lingering pattern in the residuals (errors). If there is any connection remaining in the residuals, it means that there is some pattern in the time series that the model is still unable to explain. In that circumstance, the conventional course of action is to either enhance the model's order or introduce more predictors into the system, or to seek for an alternative method to model the time series.

Checking for serial correlation ensures that the model can adequately explain the variations and patterns in the time series. The Durbin Watson's Statistic is a standard approach to check for serial correlation of errors. This statistic's value might range between 0 and 4. There is no significant serial correlation the closer it comes, approaching the value 2. The closer it is to 0, the more positive the serial correlation, and the closer it is to 4, the more negative the serial correlation. Once we have ensured that there is no autocorrelation in any of the prices, we can move on to forecasting the prices.

**Forecasting the prices**

The forecasts generated are on the scale of the training data used by the model. So, to bring it back up to its original scale, you need to de-difference it as many times as you had differenced the original input data.

*Figure 3.3: Plot of Forecast vs Actual of ICICI Bank*

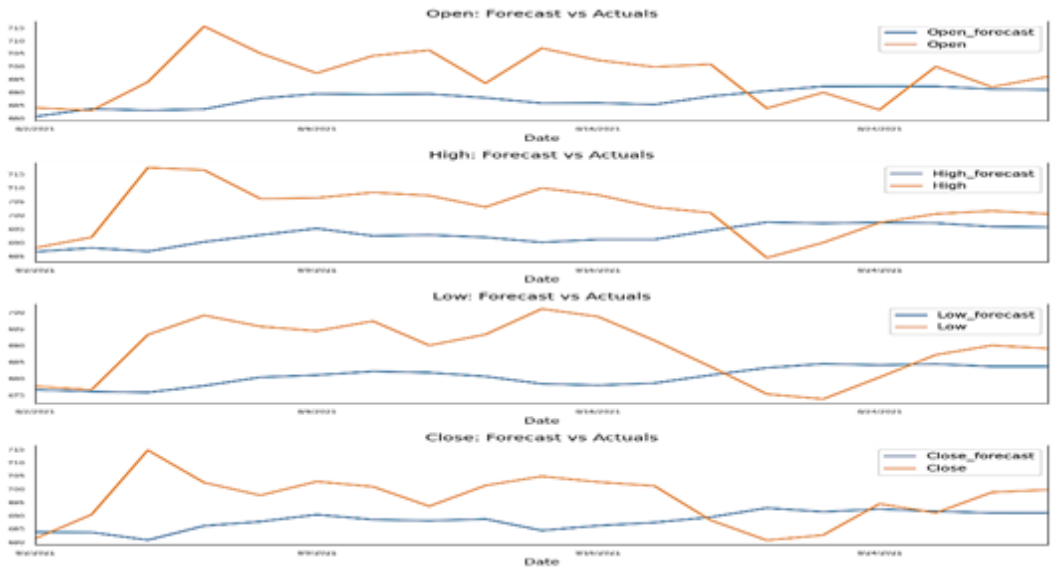

## 3.2  Sector-wise results and analysis

The performance metric used is RMSE/mean percentage. It checks what percentage of the mean of the test value is RMSE (Root Mean Square Error). RMSE/mean help to compare across stocks as the value range of the variable in consideration won't affect the metric.

### 3.2.1  Metal Sector

**Tata steel**

The graph (Fig 3.4) shows the close price of Tata steel from Jan 1, 2016 to Aug 27, 2021. It is clear from the plot that there is a sudden surge in close price during the 2020-2021 duration which supports the fact that the metal sector has outperformed during the pandemic. This also shows the importance of using walk Forward Validation (i.e., training using a small window size and recent values) instead of using the traditional train test split method.

*Figure 3.4: Plot of Tata steel close price from Jan 1, 2016 to Aug 27, 2021*

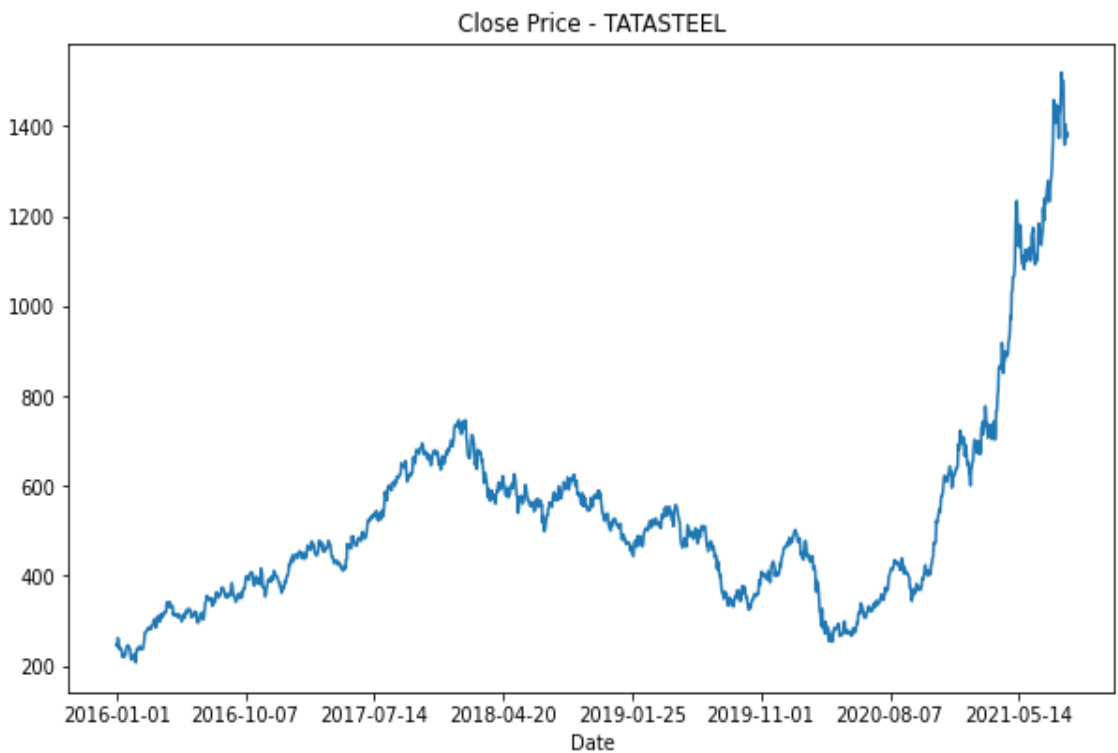

For Vector autoregression (VAR) validation is done using splitting the data into train and test. Jan 01, 2016 to Aug 10, 2021 is taken as train data and the rest 21 days are taken as the test data. The RMSE/mean obtained is 5.7815. The table below (Table 3.1) shows the comparison of results of the rest of the models

*Table 3.1: Tata steel: Comparison of expanding & sliding window validation results*

| Statistical & Econometric Models | RMSE/Mean Percentage (Expanding Window) | RMSE/Mean Percentage (Sliding Window) |
|---|---|---|
| **Linear Regression** | 3.25 | 3.80 |
| **ARIMA** | 5.60 | 5.62 |
| **MARS** | 1.08 | 1.20 |

Contrary to the belief that sliding window validation would give better results, the table shows that validation using the expanding window method has consistently given better results. Comparing the different models, MARS has given the best results followed by Linear Regression, ARIMA and VAR

**JSW Steel**

The graph (Fig 3.5) shows the close price of JSW steel from Jan 1 2016 to Aug 27, 2021. Like Tata steel, JSW steel also shows a sudden surge in close price during the 2020-2021 duration. But the range is different while the stock price of Tata steel increased from Rs. 200-400 to Rs. 400-1600 range whereas that of JSW Steel increased from Rs. 100-200 to Rs. 700-800 range.

*Figure 3.5: Plot of JSW steel close price from Jan 1, 2016 to Aug 27, 2021*

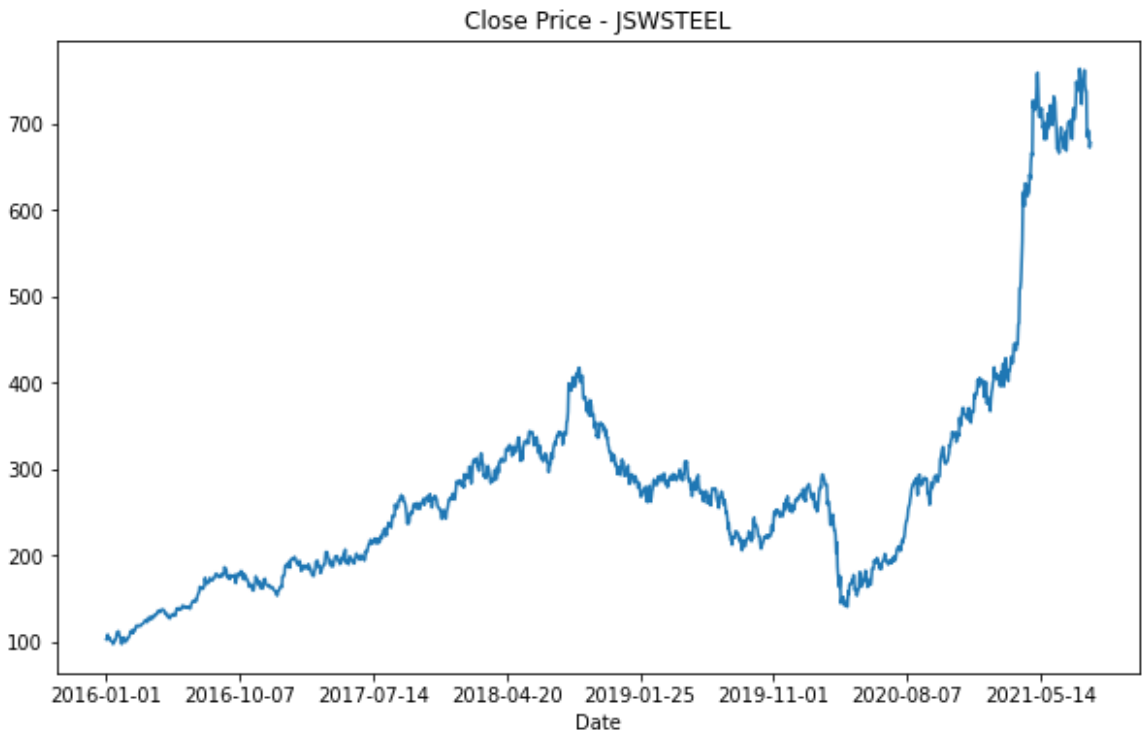

The RMSE/mean obtained for Vector autoregression(VAR) is 5.6641. The table below (Table 3.2) shows the comparison of the results of the rest of the models.

*Table 3.2: JSW steel: Comparison of Expanding & Sliding window validation results*

| Statistical & Econometric Models | RMSE/Mean Percentage (Expanding Window) | RMSE/Mean Percentage (Sliding Window) |
|---|---|---|
| Linear Regression | 13.41 | 5.42 |
| ARIMA | 5.05 | 5.04 |
| MARS | 1.07 | 1.13 |

The results show that validation using the sliding window validation method has given better results in the case of linear Regression and ARIMA and for MARS it's expanding window. Comparing the different models, MARS has given the best results followed by ARIMA, Linear Regression and VAR.

### 3.2.2 Pharma Sector

**Sun Pharma**

*Figure 3.6: Plot of Sun Pharma close price from Jan 1, 2016 to Aug 27, 2021*

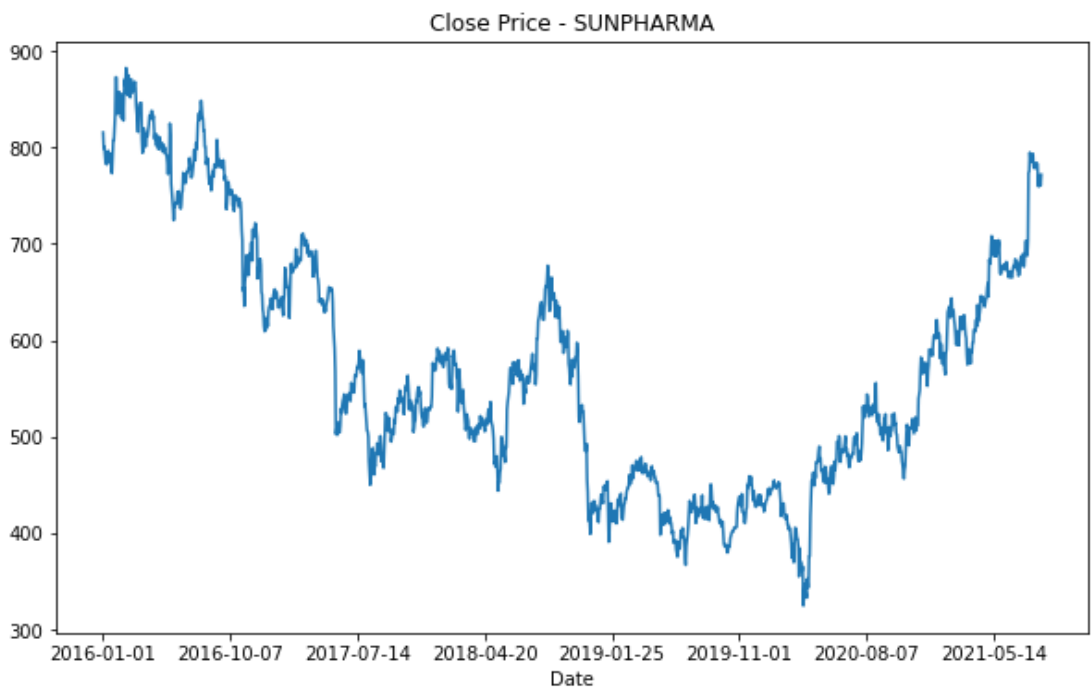

The graph (Fig 3.6) shows the close price of Sun Pharma from Jan 1, 2016 to Aug 27, 2021. The stock price was in the range of Rs 800-900 in the year 2016, it went down and came up again during the pandemic and touched the range of Rs 700-800.

The RMSE/mean obtained for Vector autoregression (VAR) is 2.8697. The table below (Table 3.3) shows the comparison of the results of the rest of the models.

*Table 3.3: Sun Pharma: Comparison of Expanding & Sliding window validation results*

| Statistical & Econometric Models | RMSE/Mean Percentage (Expanding Window) | RMSE/Mean Percentage (Sliding Window) |
|---|---|---|
| Linear Regression | 1.31 | 2.98 |
| ARIMA | 4.35 | 4.32 |
| MARS | 0.95 | 0.98 |

The results show that validation using the sliding window method has given better results in the case of ARIMA. For MARS and Linear Regression, it is the expanding window method that gives better results. Comparing the different models, MARS has given the best results followed by Linear Regression, VAR and ARIMA.

**Divi's Lab**

The graph (Fig 3.7) shows the close price of Divi's Lab from Jan 1, 2016 to Aug 27, 2021. The stock price was in the range of Rs 1000-1500 in the year 2016, it was rising and touched the range of Rs 4000-5000 in 2020-2021.

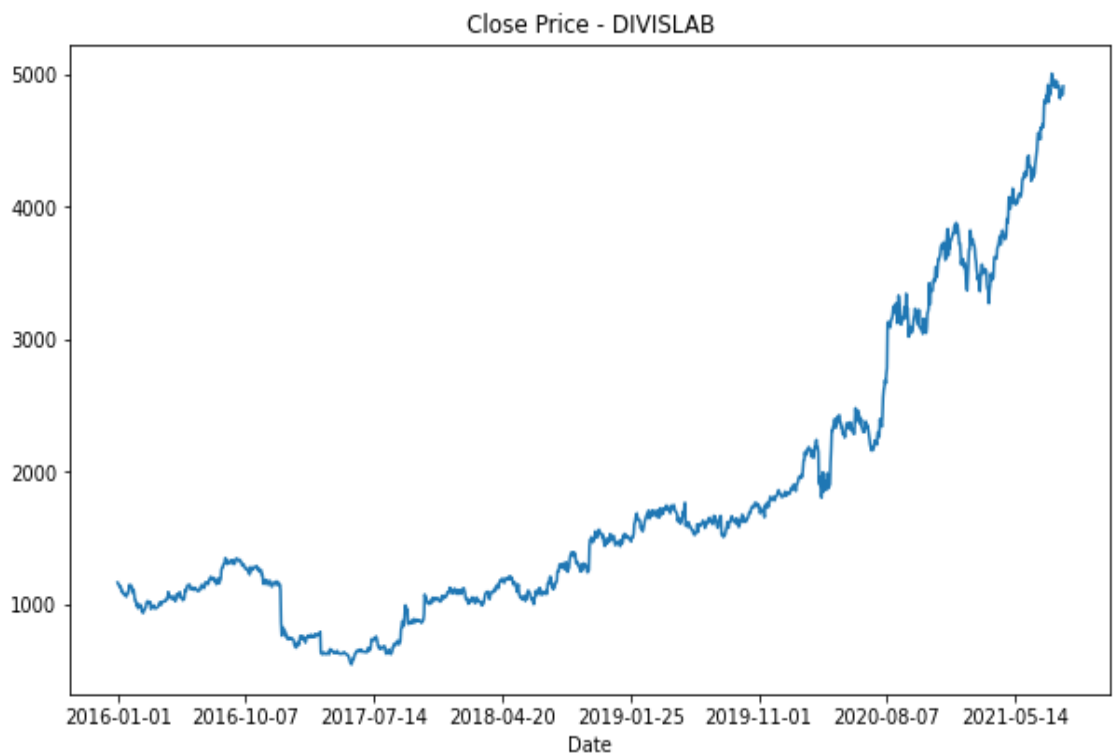

The RMSE/mean obtained for Vector autoregression(VAR) is 3.0864. The table below (Table 3.4) shows the comparison of the results of the rest of the models.

*Table 3.4: Divi's Lab: Comparison of Expanding & Sliding window validation results*

| Statistical & Econometric Models | RMSE/Mean Percentage (Expanding Window) | RMSE/Mean Percentage (Sliding Window) |
|---|---|---|
| Linear Regression | 2.55 | 4.88 |
| ARIMA | 4.45 | 4.48 |
| MARS | 0.95 | 1.15 |

The results show that validation using the expanding window method has given better results for all the models. Comparing the different models, MARS has given the best results followed by Linear Regression, VAR and ARIMA.

### 3.2.3   IT Sector

**Infosys**

The graph (Fig 3.8) shows the close price of Infosys from Jan 1 2016 to Aug 27, 2021. The stock price was in the range of Rs 500-700 in the year 2016, it was rising and touched the range of Rs 1600-1800 in 2020-2021.

*Figure 3.8: Plot of Infosys' close price from Jan 1, 2016 to Aug 27, 2021*

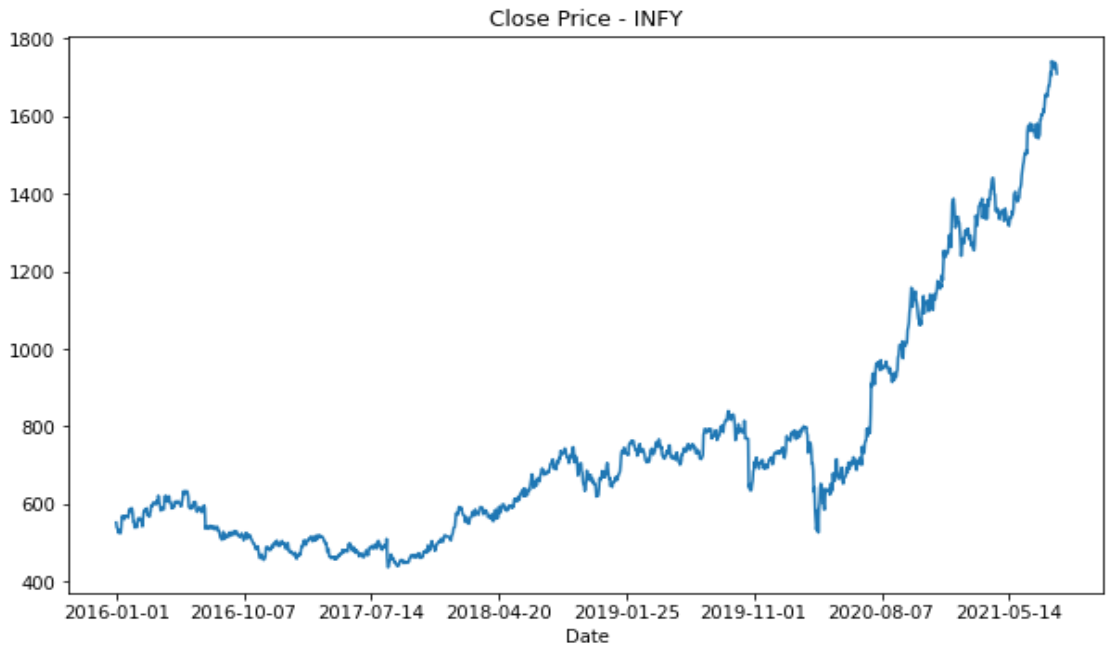

The RMSE/mean obtained for Vector autoregression(VAR) is 2.1128.  The table below (Table 3.5) shows the comparison of the results of the rest of the models.

*Table 3.5: Infosys: Comparison of Expanding & Sliding window validation results*

| Statistical & Econometric Models | RMSE/Mean Percentage (Expanding Window) | RMSE/Mean Percentage (Sliding Window) |
|---|---|---|
| Linear Regression | 1.24 | 3.31 |
| ARIMA | 3..47 | 3.47 |
| MARS | 0.76 | 0.68 |

The results show that validation using the sliding window method has given better results in the case of ARIMA and MARS. For Linear Regression it is the expanding window method that gives better results. Comparing the different models, MARS has given the best results followed by Linear Regression, VAR and ARIMA.

**TCS**

The graph (Fig 3.9) shows the close price of ITCS from Jan 1, 2016 to Aug 27, 2021. Similar to Infosys, the stock price of TCS was increasing almost steadily. The stock price was in the range of Rs 1000-1500 in the year 2016, it was rising and touched the range of Rs 3000-3500 in 2020-2021.

*Figure 3.9: Plot of TCS' close price from Jan 1, 2016 to Aug 27, 2021*

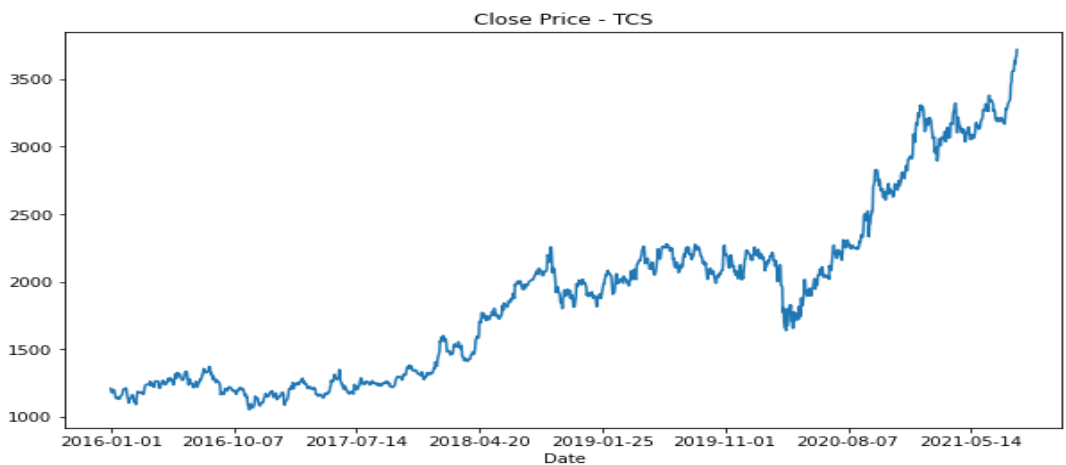

The RMSE/mean obtained for Vector autoregression(VAR) is 6.9805. The table below (Table 3.6) shows the comparison of the results of the rest of the models.

*Table 3.6: TCS: Comparison of Expanding & Sliding window validation results*

| Statistical & Econometric Models | RMSE/Mean Percentage (Expanding Window) | RMSE/Mean Percentage (Sliding Window) |
|---|---|---|
| Linear Regression | 2.05 | 4.12 |
| ARIMA | 3..43 | 3.44 |
| MARS | 1.00 | 0.69 |

The results show that validation using the sliding window method has given better results in the case of MARS. For Linear Regression and ARIMA, it is the expanding window method that gives better results. Comparing the different models, MARS has given the best results followed by Linear Regression ARIMA and VAR.

### 3.2.4 Banking Sector

**HDFC Bank**

The graph (Fig 3.10) shows the close price of HDFC Bank from Jan 1, 2016 to Aug 27, 2021. The stock price was in the range of Rs 500-600 in the year 2016, it was rising and touched the range of Rs 1400-1600 in 2020-2021.

*Figure 3.10: Plot of HDFC Bank's close price from Jan 1, 2016 to Aug 27, 2021*

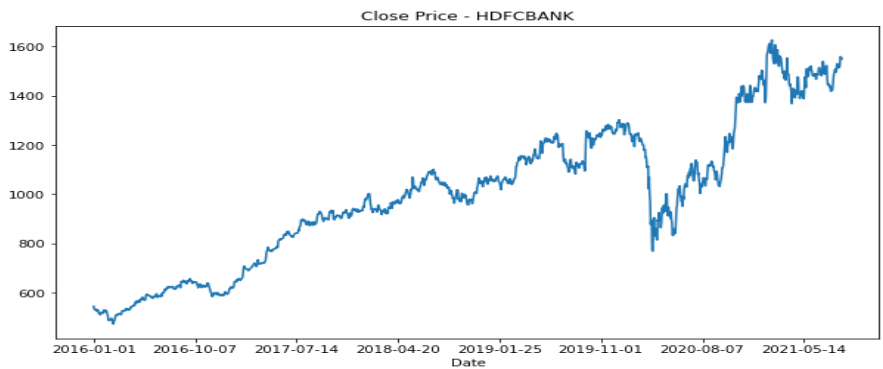

The RMSE/mean obtained for Vector autoregression (VAR) is 2.0049. The table below (Table3.7) shows the comparison of the results of the rest of the models.

*Table 3.7: HDFC Bank: Comparison of Expanding & Sliding window validation results*

| Statistical & Econometric Models | RMSE/Mean Percentage (Expanding Window) | RMSE/Mean Percentage (Sliding Window) |
|---|---|---|
| Linear Regression | 4.00 | 2.55 |
| ARIMA | 3.21 | 3.42 |
| MARS | 0.78 | 0.71 |

The results show that validation using the sliding window method has given better results in the case of Linear Regression and MARS. For ARIMA it is the expanding window method that gives better results. Comparing the different models, MARS has given the best results followed by VAR, ARIMA and Linear Regression.

**ICICI Bank**

The graph (Fig 3.11) shows the close price of ICICI Bank from Jan 1, 2016 to Aug 27, 2021. The stock price was in the range of Rs 200-300 in the year 2016, it was rising and touched the range of Rs 600-700 in 2020-2021.

*Figure 3.11: Plot of ICICI Bank's close price from Jan 1, 2016 to Aug 27, 2021*

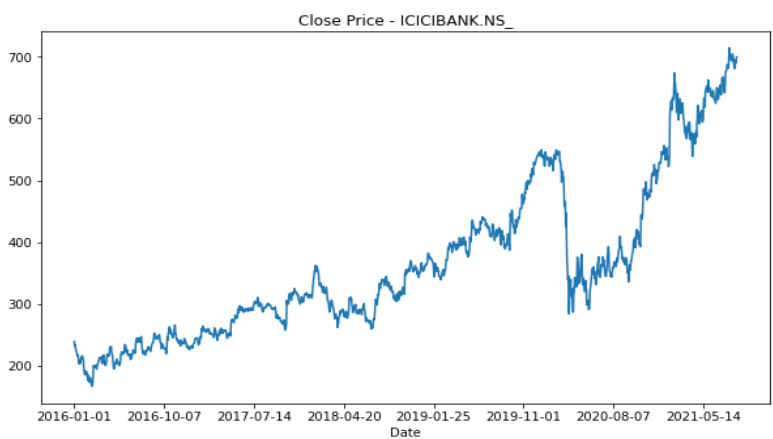

The RMSE/mean obtained for Vector autoregression (VAR) is 2.14. The table below (Table 3.8) shows the comparison of the results of the rest of the models.

*Table 3.8: ICICI Bank: Comparison of Expanding & Sliding window validation results*

| Statistical & Econometric Models | RMSE/Mean Percentage (Expanding Window) | RMSE/Mean Percentage (Sliding Window) |
|---|---|---|
| Linear Regression | 5.10 | 2.53 |
| ARIMA | 4.99 | 5.34 |
| MARS | 0.90 | 0.89 |

The results show that validation using the sliding window method has given better results in the case of Linear Regression and MARS. For ARIMA it is the expanding window method that gives better results. Comparing the different models, MARS has given the best results followed by VAR, ARIMA and Linear Regression.

### 3.2.5  Auto sector

**Maruti Suzuki**

The graph (Fig 3.12) shows the close price of Maruti Suzuki from Jan 1, 2016 to Aug 27, 2021. The stock price was in the range of Rs 4000-5000 in the year 2016, it rose and fell and again rose to the range of Rs 7000-8000 in 2020-2021.

*Figure 3.12: Plot of Maruti Suzuki's close price from Jan 1, 2016 to Aug 27, 2021*

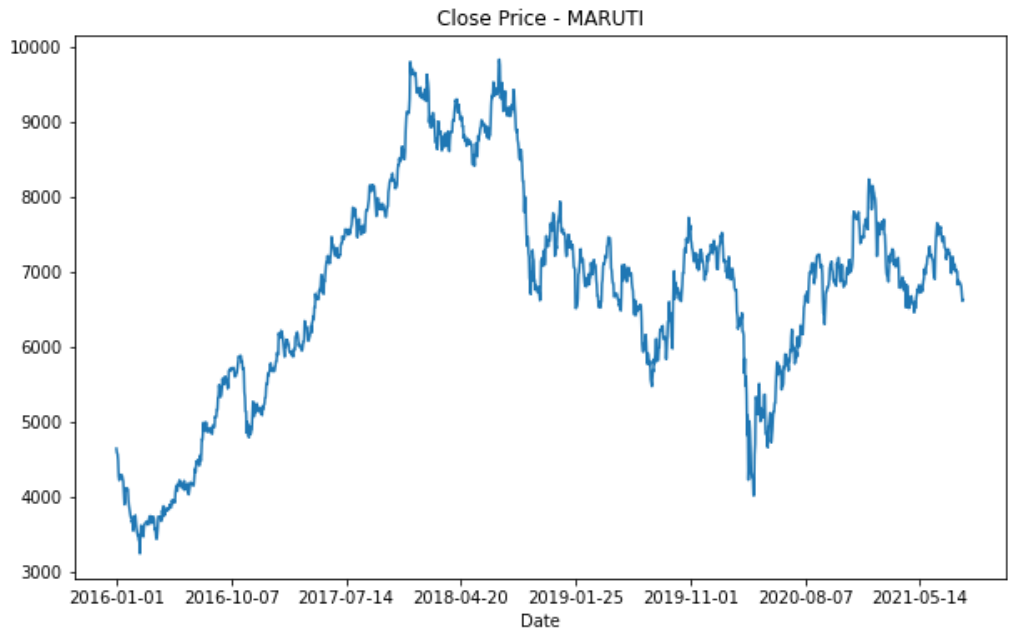

The RMSE/mean obtained for Vector autoregression (VAR) is 3.1881. The table below (Table 3.9) shows the comparison of the results of the rest of the models.

*Table 3.9: Maruti Suzuki: Comparison of Expanding & Sliding window validation results*

| Statistical & Econometric Models | RMSE/Mean Percentage (Expanding Window) | RMSE/Mean Percentage (Sliding Window) |
|---|---|---|
| Linear Regression | 0.90 | 3.41 |
| ARIMA | 4.13 | 4.16 |
| MARS | 0.83 | 3.41 |

The results show that validation using the expanding window method has given better results consistently for all three models. Comparing the different models, MARS has given the best results followed by Linear Regression, VAR and ARIMA.

**Mahindra & Mahindra**

The graph (Fig 3.13) shows the close price of Mahindra & Mahindra from Jan 1, 2016 to Aug 27, 2021. The stock price was in the range of Rs 600-800 in the year 2016, it rose and fell and again rose to the range of Rs 800-900 in 2020-2021.

*Figure 3.13: Plot of Mahindra & Mahindra's close price from Jan 1, 2016 to Aug 27, 2021*

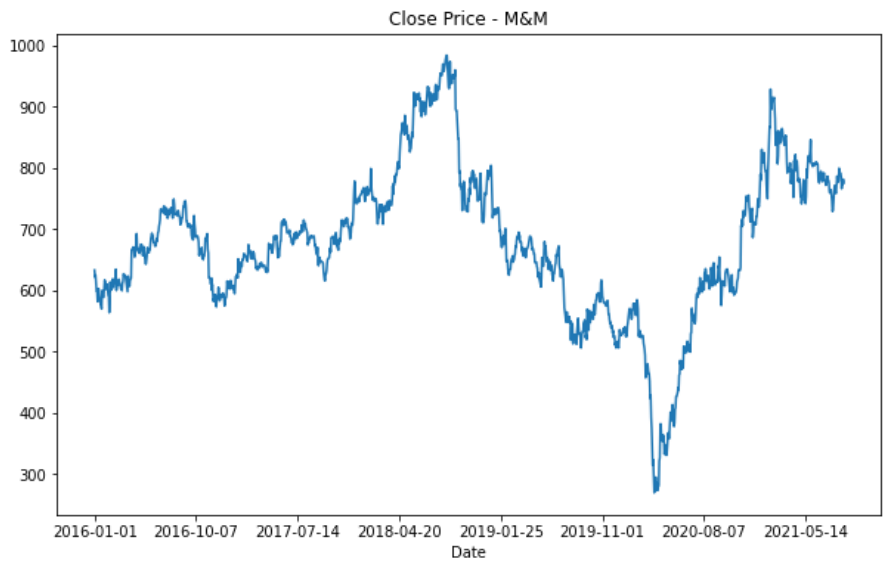

The RMSE/mean obtained for Vector autoregression (VAR) is 1.62. The table below (Table 3.10) shows the comparison of the results of the rest of the models.

*Table 3.10: Mahindra & Mahindra: Comparison of Expanding & Sliding window validation results:*

| Statistical & Econometric Models | RMSE/Mean Percentage (Expanding Window) | RMSE/Mean Percentage (Sliding Window) |
|---|---|---|
| Linear Regression | 1.34 | 3.88 |
| ARIMA | 4.41 | 4.69 |
| MARS | 0.74 | 2.61 |

The results show that validation using the expanding window method has given better results consistently for all three models. Comparing the different models, MARS has given the best results followed by Linear Regression, VAR and ARIMA.

Analyzing the performance of various econometric and statistical models in the various stocks from five sectors, it is evident that MARS (Multivariate Adaptive Regression Splines) is the model which consistently gives the best results i.e. lowest RMSE/mean value. Also sliding window and expanding window validation methods gave mixed results, which makes it difficult to conclude that one method is better than the other.

# Chapter 4

## 3.3   Machine Learning Models

Five machine learning models are used for Regression (K Nearest Neighbor, Decision Tree, Random Forest, XGBoost, SVM) and six machine learning models are used for classification (Logistic Regression along with other models used for regression). Each of them is explained in detail in this chapter. For regression, the predictor and the target variables used are as mentioned in Chapter 2. For classification the predictor variables and target variables used are as follows:

**Predictors**

    I.    Day of the week: 0-4, represents days from Monday to Friday in order

    II.    Day of the month: 1-31, represents the 31 days of a month

    III.    Month: 1-12, represents the 12 months of a year

    IV.    Open_Perc: Percentage change in open price w.r.t the previous day's open price

    V.    High_Perc: Percentage change in High value of close price w.r.t the previous day

    VI.    Low_Perc: Percentage change in Low value of close price w.r.t the previous day

    VII.    NIFTY50_perc: Percentage change in the NIFTY50 index w.r.t that of the previous day.

VIII.   Volume_perc: Percentage change in the no of stocks traded w.r.t that of the previous day.

**Target**

Close_perc_label:  If the percentage change in close price w.r.t the previous day is 0 or positive, then the value is 1 and if it is negative then the value is 0.

The walk-forward validation method is used for model validation of ML classification and regression models. Hyperparameter tuning is not done for any of the models since the training and testing are happening in an iterative manner for a small window size (test data= 14 data points, train data (sliding window) =245 data points).

### 3.3.1  K Nearest Neighbor

K nearest neighbor is the simplest machine learning model which works based on the logic of proximity. During the training phase, the hyperparameters like k value and distance metric are learned and it is during the testing phase, that the calculations to find the nearest neighbor happen. Since hyperparameters were not tuned, the default hyperparameters used are k=5, distance metric = euclidean distance, and weight='uniform' (equal weightage given for all the neighbors). The value of $k$ determines how many closest similar records in the training data set are considered for classification or regression.

### 3.3.2  Decision Tree

When a dataset needs to be divided into classes that correspond to the response variable, classification trees are used. The classes 'Yes' or 'No' are frequently used. In other words, there are only two of them, and they are mutually exclusive. When the response variable is continuous, however, regression trees are used. A regression tree is employed, for example, if the response variable is the price of a property or the temperature of the day. To put it another way, regression trees are used to solve prediction problems, whereas classification trees are used to solve classification problems.

To control the tree depending on impurity levels, use the **min impurity split** argument. It imposes a limit on gini. If the min impurity split is set to 0.3, a node must have a gini value greater than 0.3 in order to be further splitted. **max depth** is another hyperparameter for controlling the depth of a tree. It does not do any impurity or sample ratio calculations. When max depth is achieved, the model stops dividing.

### 3.3.3 Support Vector Machine

A support vector machine (SVM) is a classification and regression machine learning model. It can categorize both linear and nonlinear data when used for classification. The original training data is transformed into a higher dimension via a nonlinear mapping. It looks for the linear ideal hyperplane that separates the two classes inside this new higher dimension. Support vectors, which are the essential and discriminating training tuples to distinguish the two classes, are used by SVM to find this hyperplane.

### 3.3.4 Random Forest

Random forest is a supervised machine learning algorithm that is commonly used to solve classification and regression problems. It creates decision trees from various samples, using the majority vote for classification and the average for regression. One of the most essential characteristics of the Random Forest Algorithm is that it can handle data sets with both continuous and categorical variables, as in regression and classification. When it comes to categorization difficulties, it outperforms the competition.

### 3.3.5 **XGBoost**

Boosting is nothing but ensemble techniques where previous model errors are resolved in the new models. These models are added straight until no other improvement is seen. One of the best examples of such an algorithm is the AdaBoost algorithm. Gradient boosting is a method where the new models are created that computes the error in the previous model and then leftovers are added to make the final prediction.

### 3.3.6 **Logistic Regression**

Logistic Regression is a Machine Learning algorithm that is used for classification problems; it is a predictive analysis algorithm based on the concept of probability. A Logistic Regression model is similar to a Linear Regression model, except that the Logistic Regression uses a more complex cost function, which is known as the 'Sigmoid function' or the 'logistic function' instead of a linear function.

## 3.4 **Sector-wise results analysis for ML classification and Regression**

The performance metric used for regression is RMSE/mean percentage and for classification, it is accuracy percentage.

### 3.4.1 **Metal Sector**

**Tata Steel**

A comparison of the performance of all the regression models and classification models using expanding window validation method and sliding window validation method is done

**Regression results:**

*Table 0.1: Tata Steel: Expanding & Sliding window validation results for ML Regression models*

| ML Regression Models | RMSE/Mean Percentage (Expanding Window) | RMSE/Mean Percentage (Sliding Window) |
|---|---|---|
| **Decision Tree** | 2.24 | 3.49 |
| **KNN** | 6.51 | 6.51 |
| **Random Forest** | 2.03 | 3.03 |
| **SVM** | 26.71 | 16.83 |
| **XG Boost** | 2.15 | 3.15 |

The results show that the expanding window validation method gave better results for Decision Tree, Random Forest, and XGBoost whereas the sliding window method gave better results for Random Forest and XGBoost. For KNN, both expanding window and sliding window methods gave the same results. Among the models Random Forest performed the best with the lowest RMSE/ mean value.

'Day of the week' plays an important role in the stock price prediction. Prediction on a Monday wouldn't be equally accurate as of the prediction in the middle of the week or on a Friday. To understand the day-wise RMSE/mean behavior, RMSE/mean values for each of the weekday is segregated and the mean of the same is found and has been plotted.

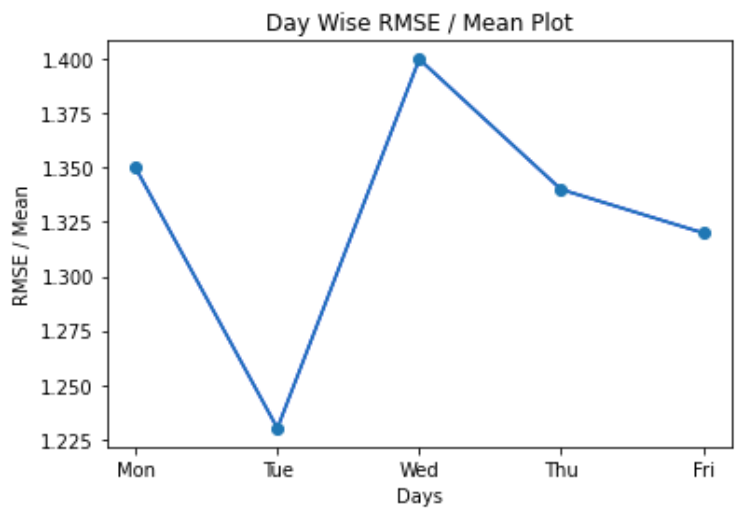

The plot shows that the RMSE/mean value is highest for Wednesday and lowest on Tuesday.

**Classification results:**

*Table 0.2: Tata Steel: Expanding & Sliding window validation results for Classification models*

| ML Classification Models | Accuracy (Expanding Window) | Accuracy (Sliding Window) |
|---|---|---|
| **Decision Tree** | 77.57 | 73.30 |
| **KNN** | 51.59 | 48.91 |
| **Random Forest** | 83.24 | 79.67 |
| **SVM** | 54.39 | 51.74 |
| **XG Boost** | 83.50 | 79.16 |
| **Logistic Regression** | 62.46 | 54.39 |

The results show that the expanding window validation method gave better accuracy consistently for all the models.

## JSW Steel

## Regression results:

*Table 0.3: JSW Steel: Expanding & Sliding window validation results for ML Regression models*

| ML Regression Models | RMSE/Mean Percentage (Expanding Window) | RMSE/Mean Percentage (Sliding Window) |
|---|---|---|
| Decision Tree | 2.43 | 3.09 |
| KNN | 6.18 | 6.18 |
| Random Forest | 2.24 | 3.00 |
| SVM | 28.65 | 16.24 |
| XG Boost | 2.35 | 3.09 |

The results show that the expanding window validation method gave better results for Random Forest, XGBoost, and Decision Tree. whereas the sliding window method gave better results for Random Forest. For KNN, both expanding window and sliding window methods gave the same results.

## Classification results:

*Table 0.4: JSW Steel: Expanding & Sliding window validation results for Classification models*

| ML Classification Models | Accuracy (Expanding Window) | Accuracy (Sliding Window) |
|---|---|---|
| Decision Tree | 73.74 | 69.39 |
| KNN | 52.21 | 51.71 |
| Random Forest | 79.38 | 75.40 |
| SVM | 55.79 | 53.14 |
| XG Boost | 80.37 | 76.32 |
| Logistic Regression | 58.37 | 53.00 |

The results show that the expanding window validation method gave better accuracy consistently for all the models.

### 3.4.2  Pharma Sector

**Sun Pharma**

**Regression results:**

*Table 0.5: Sun Pharma: Expanding & Sliding window validation results for ML Regression models*

| ML Regression Models | RMSE/Mean Percentage (Expanding Window) | RMSE/Mean Percentage (Sliding Window) |
|---|---|---|
| **Decision Tree** | 1.87 | 2.58 |
| **KNN** | 5.23 | 5.23 |
| **Random Forest** | 1.52 | 2.18 |
| **SVM** | 21.87 | 9.77 |
| **XG Boost** | 1.63 | 2.37 |

The results show that the expanding window validation method gave better results for Random Forest, XGBoost and Decision Tree. whereas the sliding window method gave better results for Random Forest. For KNN, both expanding window and sliding window methods gave the same results.

*Figure 0.2: Sun Pharma: Day-wise RMSE/Mean plot for ML model*

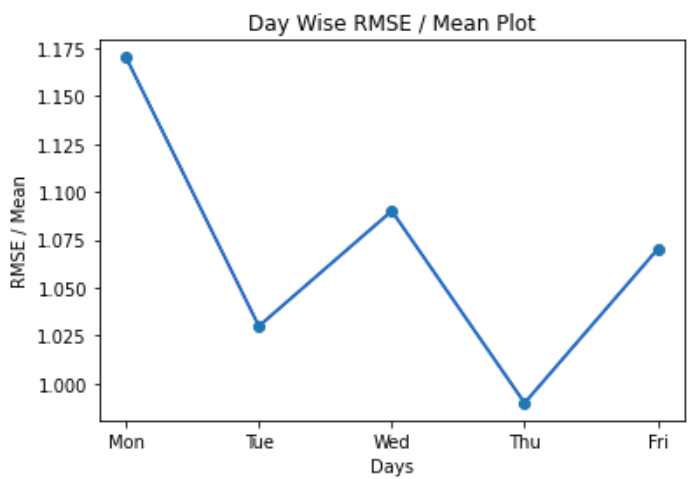

The plot shows that the prediction is more accurate on Thursdays and Tuesdays compared to other days as the RMSE/mean value is lower for those days.

**Classification results:**

*Table 0.6: Sun Pharma: Expanding & Sliding window validation results for Classification models*

| ML Classification Models | Accuracy (Expanding Window) | Accuracy (Sliding Window) |
|---|---|---|
| **Decision Tree** | 72.73 | 73.52 |
| **KNN** | 49.81 | 49.60 |
| **Random Forest** | 81.09 | 77.41 |
| **SVM** | 51.19 | 48.98 |
| **XG Boost** | 80.10 | 77.96 |
| **Logistic Regression** | 60.99 | 52.04 |

The results show that the expanding window validation method gave better accuracy than the sliding window. As we can observe that the Random Forest model gives high accuracy among others.

**Divi's Lab**

**Regression results:**

*Table 0.7: Divi's Lab: Expanding & Sliding window validation results for ML Regression models*

| ML Regression Models | RMSE/Mean Percentage (Expanding Window) | RMSE/Mean Percentage (Sliding Window) |
|---|---|---|
| Decision Tree | 2.57 | 3.07 |
| KNN | 5.79 | 5.79 |
| Random Forest | 2.31 | 2.73 |
| SVM | 37.44 | 14.76 |
| XG Boost | 2.33 | 2.69 |

The results show that the expanding window validation gave better results for Random Forest, XGBoost and Decision Tree. whereas the sliding window method gave better results for XGBoost. For KNN, both methods gave the same results.

**Classification results:**

*Table 0.8: Divi's Lab: Expanding & Sliding window validation results for Classification models*

| ML Classification Models | Accuracy (Expanding Window) | Accuracy (Sliding Window) |
|---|---|---|
| Decision Tree | 71.60 | 69.94 |
| KNN | 51.38 | 50.16 |
| Random Forest | 77.81 | 75.29 |
| SVM | 54.40 | 52.85 |
| XG Boost | 77.78 | 74.58 |
| Logistic Regression | 60.11 | 53.07 |

The results show that the expanding window validation method gave better accuracy than the sliding window. As we can observe that the Random Forest model gives high accuracy among others.

### 3.4.3 IT Sector

**Infosys**

**Regression results:**

*Table 0.9: Infosys: Expanding & Sliding window validation results for ML Regression models*

| ML Regression Models | RMSE/Mean Percentage (Expanding Window) | RMSE/Mean Percentage (Sliding Window) |
|---|---|---|
| **Decision Tree** | 1.61 | 1.92 |
| **KNN** | 4.23 | 4.23 |
| **Random Forest** | 1.46 | 1.85 |
| **SVM** | 23.28 | 9.35 |
| **XG Boost** | 1.50 | 1.97 |

The results show that the expanding window validation method gave better results for Random Forest, XGBoost and Decision Tree. whereas the sliding window method gave better results for Random Forest. For KNN, both expanding window and sliding window methods gave the same results.

*Figure 0.3: Infosys: Day-wise RMSE/Mean plot for ML model*

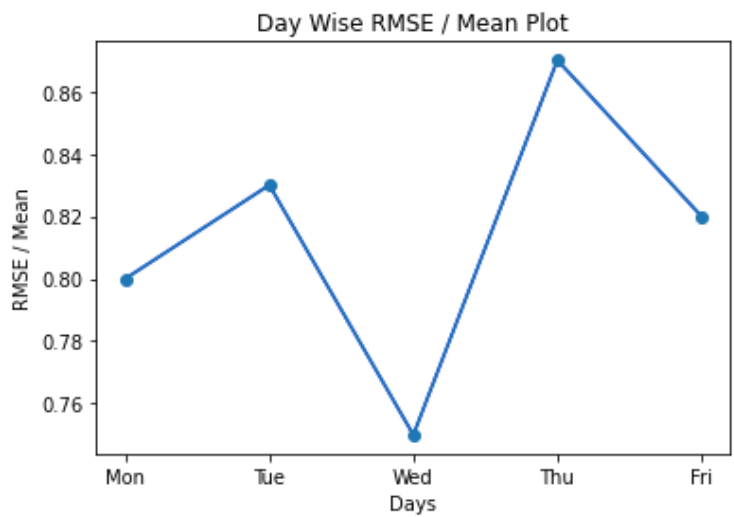

The plot shows that the prediction is more accurate on Wednesday compared to other days as the RMSE/mean value is the lowest on Wednesday.

**Classification results:**

*Table 0.10: Infosys: Expanding & Sliding window validation results for Classification models*

| ML Classification Models | Accuracy (Expanding Window) | Accuracy (Sliding Window) |
|---|---|---|
| **Decision Tree** | 74.77 | 70.71 |
| **KNN** | 51.71 | 51.96 |
| **Random Forest** | 81.77 | 78.34 |
| **SVM** | 55.09 | 54.58 |
| **XG Boost** | 82.50 | 78.97 |
| **Logistic Regression** | 58.55 | 51.41 |

The results show that the expanding window validation method gave better accuracy than the sliding window. As we can observe that the XGBoost model gives high accuracy among others.

## TCS

## Regression results:

*Table 0.11: TCS: Expanding & Sliding window validation results for ML Regression models*

| ML Regression Models | RMSE/Mean Percentage (Expanding Window) | RMSE/Mean Percentage (Sliding Window) |
|---|---|---|
| Decision Tree | 1.72 | 1.87 |
| KNN | 4.24 | 4.24 |
| Random Forest | 1.44 | 1.78 |
| SVM | 25.37 | 7.94 |
| XG Boost | 1.53 | 1.86 |

The results show that the expanding window validation method gave better results for Random Forest, XGBoost and Decision Tree. whereas the sliding window method gave better results for Random Forest. For KNN, both expanding window and sliding window methods gave the same results.

## Classification results:

*Table 0.12: TCS: Expanding & Sliding window validation results for Classification models*

| ML Classification Models | Accuracy (Expanding Window) | Accuracy (Sliding Window) |
|---|---|---|
| Decision Tree | 74.21 | 73.99 |
| KNN | 50.45 | 52.97 |
| Random Forest | 79.92 | 78.45 |
| SVM | 54.70 | 54.70 |
| XG Boost | 82.43 | 79.48 |
| Logistic Regression | 57.83 | 54.46 |

The results show that the expanding window validation method gave better accuracy than the sliding window. As we can observe that the XGBoost model gives high accuracy among others.

3.4.4   **Banking Sector**

**HDFC Bank**

**Regression results:**

*Table 0.13: HDFC Bank: Expanding & Sliding window validation results for ML Regression models*

| ML Regression Models | RMSE/Mean Percentage (Expanding Window) | RMSE/Mean Percentage (Sliding Window) |
|---|---|---|
| **Decision Tree** | 1.53 | 1.86 |
| **KNN** | 3.51 | 3.51 |
| **Random Forest** | 1.28 | 1.59 |
| **SVM** | 21.03 | 8.36 |
| **XG Boost** | 1.41 | 1.80 |

The results show that the expanding window validation method gave better results for Random Forest, XGBoost and Decision Tree. whereas the sliding window method gave better results for Random Forest. For KNN, both expanding window and sliding window methods gave the same results.

*Figure 0.4: HDFC Bank: Day-wise RMSE/Mean plot for ML model*

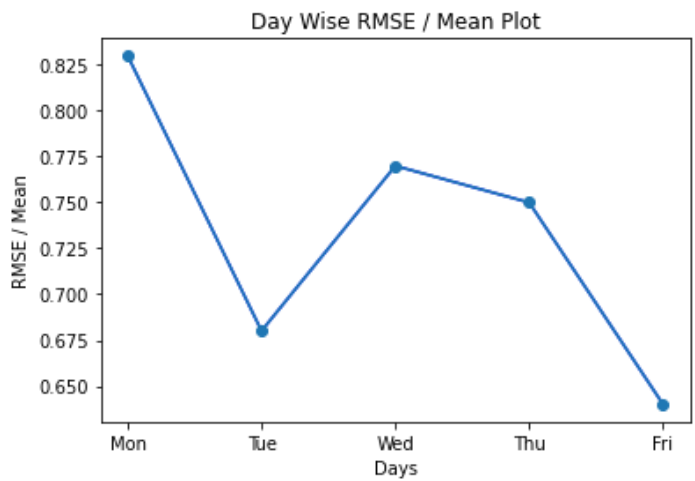

The plot shows that the prediction is more accurate on Friday compared to other days as the RMSE/mean value is the lowest on Friday.

**Classification results:**

*Table 0.14: HDFC Bank: Expanding & Sliding window validation results for Classification models*

| ML Classification Models | Accuracy (Expanding Window) | Accuracy (Sliding Window) |
|---|---|---|
| **Decision Tree** | 74.80 | 73.40 |
| **KNN** | 52.30 | 49.54 |
| **Random Forest** | 80.46 | 76.89 |
| **SVM** | 55.09 | 52.85 |
| **XG Boost** | 80.98 | 77.52 |
| **Logistic Regression** | 56.42 | 52.37 |

The results show that the expanding window validation method gave better accuracy than the sliding window. As we can observe that the XGBoost model gives high accuracy among others.

## ICICI Bank

**Regression results:**

*Table 0.15: ICICI Bank: Expanding & Sliding window validation results for ML Regression models*

| ML Regression Models | RMSE/Mean Percentage (Expanding Window) | RMSE/Mean Percentage (Sliding Window) |
|---|---|---|
| Decision Tree | 2.08 | 2.42 |
| KNN | 5.45 | 5.45 |
| Random Forest | 1.64 | 2.15 |
| SVM | 23.98 | 11.36 |
| XG Boost | 1.81 | 2.31 |

The results show that the expanding window validation method gave better results for Random Forest, XGBoost, and Decision Tree. whereas the sliding window method gave better results for Random Forest. For KNN, both expanding window and sliding window methods gave the same results.

**Classification results:**

*Table 0.16: ICICI Bank: Expanding & Sliding window validation results for Classification models*

| ML Classification Models | Accuracy (Expanding Window) | Accuracy (Sliding Window) |
|---|---|---|
| Decision Tree | 76.09 | 73.43 |
| KNN | 52.48 | 51.83 |
| Random Forest | 81.91 | 79.07 |
| SVM | 51.45 | 50.46 |
| XG Boost | 83.24 | 79.92 |
| Logistic Regression | 60.99 | 51.08 |

The results show that the expanding window validation method gave better accuracy than the sliding window. As we can observe that the XGBoost model gives high accuracy among others.

### 3.4.5 **Auto Sector**

**Maruti Suzuki**

**Regression results:**

*Table 0.17: Maruti Suzuki: Expanding & Sliding window validation results for ML Regression models*

| ML Regression Models | RMSE/Mean Percentage (Expanding Window) | RMSE/Mean Percentage (Sliding Window) |
|---|---|---|
| **Decision Tree** | 1.62 | 2.42 |
| **KNN** | 4.95 | 4.95 |
| **Random Forest** | 1.38 | 2.16 |
| **SVM** | 17.24 | 10.95 |
| **XG Boost** | 1.51 | 2.30 |

The results show that the expanding window validation method gave better results for Random Forest, XGBoost, and Decision Tree. whereas the sliding window method gave better results for XGBoost. For KNN, both expanding window and sliding window methods gave the same results.

*Figure 0.5: Maruti Suzuki: Day-wise RMSE/Mean plot for ML model*

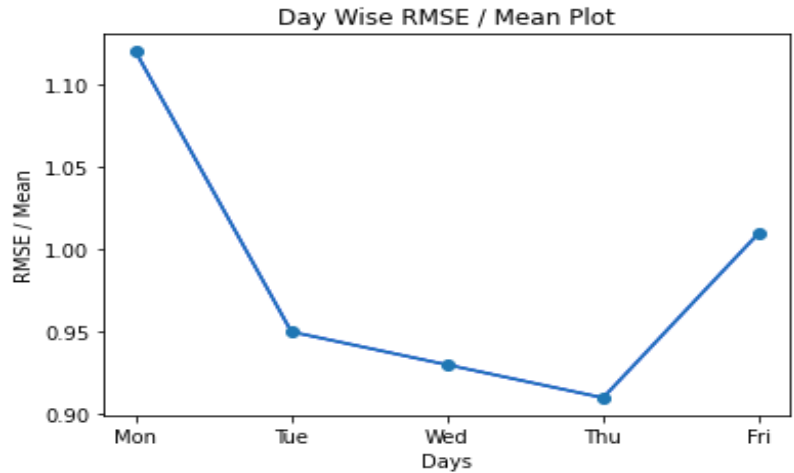

The plot shows that the prediction is more accurate on Thursday compared to other days as the RMSE/mean value is lower.

**Classification results:**

*Table 0.18: Maruti Suzuki: Expanding & Sliding window validation results for Classification models*

| ML Classification Models | Accuracy (Expanding Window) | Accuracy (Sliding Window) |
|---|---|---|
| **Decision Tree** | 74.80 | 74.31 |
| **KNN** | 50.01 | 50.74 |
| **Random Forest** | 80.84 | 78.89 |
| **SVM** | 53.33 | 51.67 |
| **XG Boost** | 81.25 | 80.33 |
| **Logistic Regression** | 56.79 | 52.70 |

The results show that the expanding window validation method gave better accuracy than the sliding window. As we can observe that the XGBoost model gives high accuracy among others.

## Mahindra & Mahindra

## Regression results:

*Table 0.19: Mahindra & Mahindra: Expanding & Sliding window validation results for ML Regression models*

| ML Regression Models | RMSE/Mean Percentage (Expanding Window) | RMSE/Mean Percentage (Sliding Window) |
|---|---|---|
| Decision Tree | 1.87 | 2.63 |
| KNN | 5.13 | 5.13 |
| Random Forest | 1.65 | 2.42 |
| SVM | 16.24 | 11.64 |
| XG Boost | 1.76 | 2.50 |

The results show that the expanding window validation method gave better results for Random Forest, XGBoost and Decision Tree. whereas the sliding window method gave better results for XGBoost. For KNN, both expanding window and sliding window methods gave the same results.

## Classification results:

*Table 0.20: Mahindra & Mahindra: Expanding & Sliding window validation results for Classification models*

| ML Classification Models | Accuracy (Expanding Window) | Accuracy (Sliding Window) |
|---|---|---|
| Decision Tree | 75.14 | 71.19 |
| KNN | 52.90 | 48.60 |
| Random Forest | 80.59 | 77.63 |
| SVM | 52.00 | 51.30 |
| XG Boost | 81.03 | 77.71 |
| Logistic Regression | 60.47 | 52.70 |

The results show that the expanding window validation method gave better accuracy than the sliding window. As we can observe that the XGBoost model gives high accuracy among others.

Analyzing the performance of various ML models in multiple stocks of 5 different sectors, it could be noticed that Random Forest and XGBoost are the two models that performed better than all other models both in the case of regression and classification. Also, between sliding window validation method and expanding window method, there is no clear winner. While for some of the stocks and models expanding window performs better, n some other cases it's the sliding window that is performing better.

4 **Chapter 5**

## 4.1 **Deep Learning Models**

Two deep learning regression models are used: (i) the long- and short-term memory (LSTM) network, and (ii) the convolutional neural networks (CNNs). The same along with the performance of the models on 10 different stocks is explained in detail in this chapter.

### 4.1.1 **Long- and Short-Term Memory Network**

LSTM is a different type of normal neural network (RNN) - neural network with return/feedback loops. In such networks, current output depends on the current input and previous network status. However, RNNs suffer from the problem that these networks are unable to capture long-term dependence due to vanishing or exploding gradients during backpropagation. LSTM networks overcome such problems, so such networks work well in predicting univariate time series. LSTM networks contain memory cells that are able to retain their state over time using memory and input units that control the entry and exit of information from memory. There are different types of gates used. Forget gates control what information to be removed from memory. Input gates are designed to control new information that can be added to the cell state from the current input. The cell state vector includes two components - the old memory from the forget gate, and the new memory from the input gate. Finally, the output gates determine in terms of what they will extract from the memory cells. The LSTM network architecture and backpropagation through time (BPTT) learning algorithm provides such networks with a powerful ability to learn and predict for a univariate time series. We use Python programming language and Tensorflow learning framework to launch LSTM networks and use those networks to predict stock prices. The performance of two different LSTM networks are studied, one with an input of the past 5 days' Close price and predicting the next 5 days' Close Price (Fig 21) i.e. input as 5 data points and output as 5 data points.  In the second case, input is past 10 days' Close price and predicted the next 5 days' Close Price (Fig 22). The performance of the models is also studied by varying the number of nodes in the LSTM layer. The objective of the same is to analyze how the increase or decrease in parameters would affect the accuracy of prediction and reduce/increase the time taken for computation. The loss function used is 'mean absolute error '(MAE), the optimizer used is Adam optimizer, and the activation function used is ReLU (Rectified Linear Unit).

Out of the 1745 records, the first 1000 records are used for training, and the remaining 745 for the validation. A batch size of 4 and an epoch value of 20 is used. The Sequential function defined in Keras is used for building the LSTM and the model is compiled using MAE as the loss function and ADAM as the optimizer. The model architecture with an input of N=5 is depicted in Fig, (21). The input layer consists of a single time series data with 5 values and the output of the input layer is passed on to the LSTM layer with 200 nodes. The output of the LSTM layer is passed on to a dense layer (i.e., a fully connected layer) that has 100 nodes and is further connected to a dense layer that has 5 nodes. From the final output layer, we get the output of N=5. The training and validation losses are found to have converged to a low value.

*Figure 4.1: LSTM model architecture – 5 days' data as input (N = 5) and 5 days' data output*

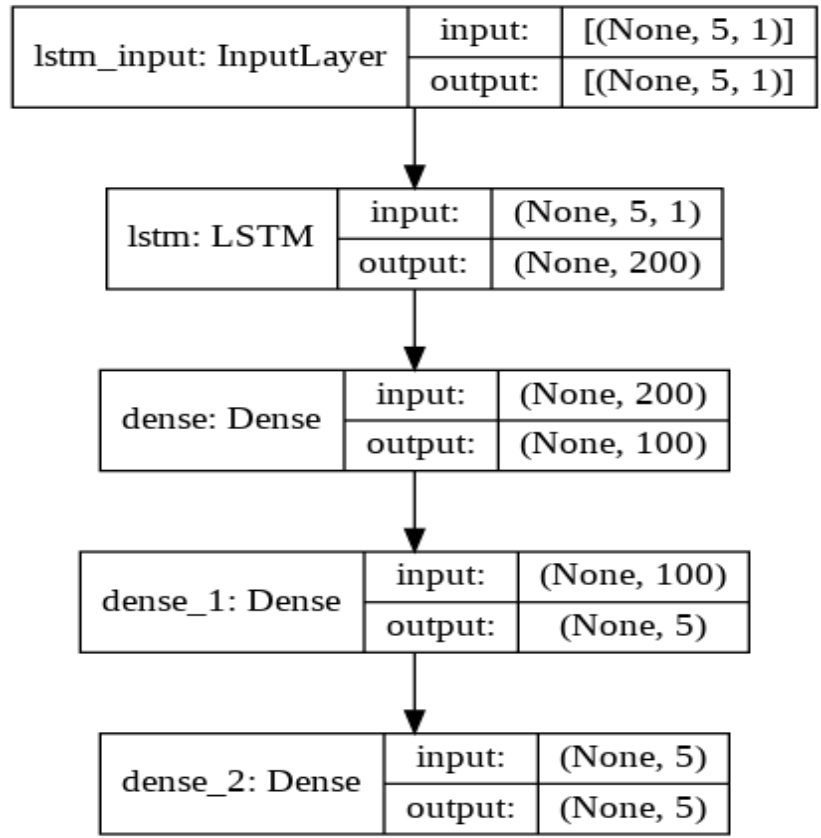

The model architecture with an input of N=10 is depicted in Fig, (22). The input layer consists of a single time series data with 10 values and the output of the input layer is passed on to the LSTM layer with 200 nodes. The output of the LSTM layer is passed on to a dense layer (i.e., a fully connected layer) that has 100 nodes and is further connected to a dense layer that has 5 nodes. From the final output layer, we get the output of N=5. The training and validation losses are found to have converged to a low value.

*Figure 4.2: LSTM model architecture – 10 days' data as input (N = 10) and 5 days' data output*

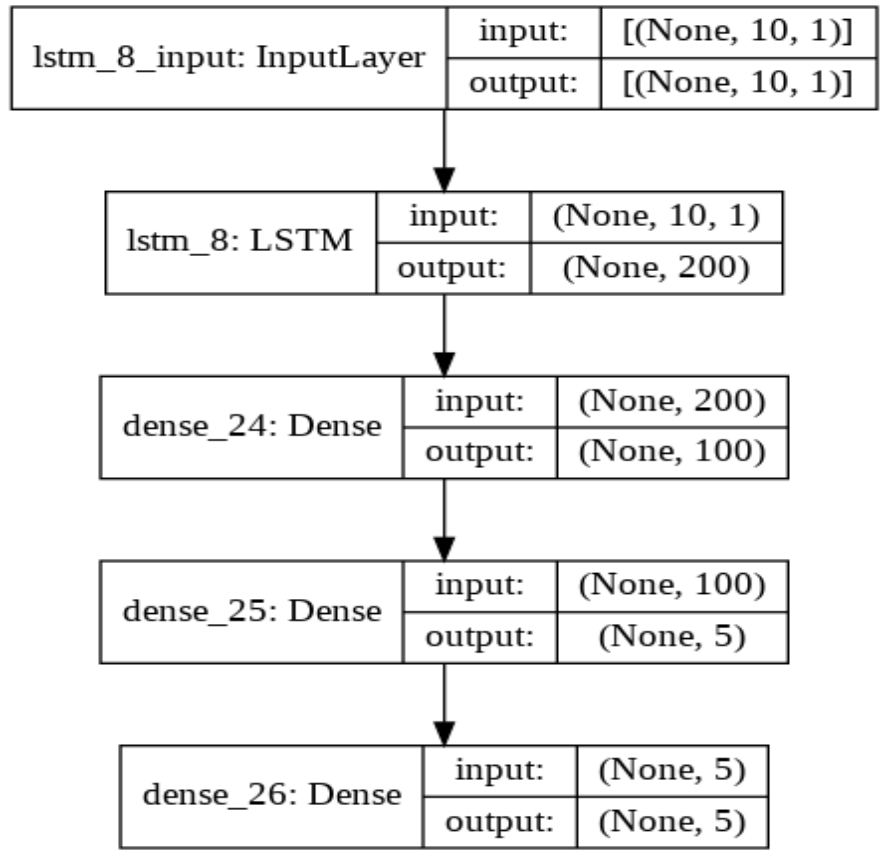

### 4.1.2 Convolutional Neural Networks

CNN consists of a series of convolutional layers. The output of the convolutional layers is connected only to the local regions in the input. This is obtained by sliding the filter, or the weight matrix, over the input, and at each point computing the dot product between the weight matrix and the input region. This structure allows the model to read filters that can detect certain patterns in the input data.

In the present work, we have used CNN to forecast the univariate time series data. CNN has two important processing layers, convolutional layers, and pooling layers. The convolutional layers read input using a filter by scanning across the input data field. The output of the convolutional layer is an interpretation of the input that is projected onto the filter map. The pooling layer takes the projections and reduces them to the most essential elements, using average pool or max pool. The convolution and pooling layers are repeated and the output of the final pooling layer is provided to one or more fully-connected layers that interpret what has been read.

The performance of two different CNNs is studied, one with an input of the past 5 days' Close price' and predicting the next 5 days 'Close Price' (Fig 21) i.e. input as 5 data points and output as 5 data points.  In the second case, input is past 10 days' Close price and predicted the next 5 days' Close Price (Fig 22). The performance of the models is also studied by varying the number of filters. The objective of the same is to analyze how the increase or decrease in parameters would affect the accuracy of prediction and reduce/increase the time taken for computation. The loss function used is 'mean absolute error '(MAE), the optimizer used is Adam optimizer, and the activation function used is ReLU (Rectified Linear Unit).

The model architecture with an input of N=5 is depicted in Fig, (23). Only one convolution layer has been used with 16 filters and a kernel size of 3. In other words, it means that the input sequence of five days is read with a convolutional operation in three time-steps at a time and this operation is performed 16 times. A max-pooling layer of size 2 is used that reduces the size of the feature maps into half before the internal representation is flattened to one long vector. This is further connected to a fully-connected layer of 10 nodes before the output layer with 5 nodes (which is also fully connected) that predicts the close price for the next five days.

*Figure 4.3: CNN model architecture – 5 days' data as input (N = 5) and  5 days' data output*

**64**

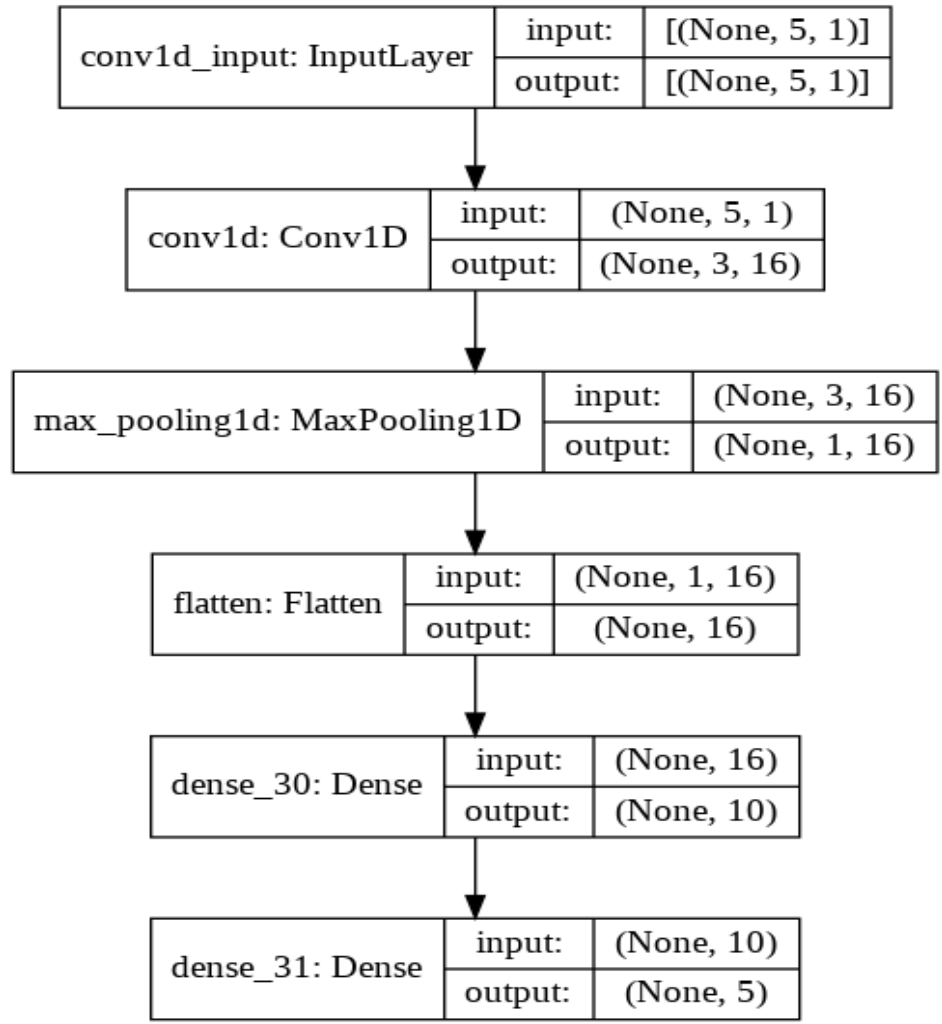

 For an input of N=10 also a similar model architecture Fig, (23) has been used. Here also, we used only one convolution layer with 16 filters and a kernel size of 3. A max-pooling layer of size 2 is used that reduces the size of the feature maps into half before the internal representation is flattened to one long vector. This is further connected to a fully-connected layer of 10 nodes before the output layer with 5 nodes (which is also fully connected) ) that predicts the close price for the next five days.

*Figure 4.4: CNN model architecture – 10 days' data as input (N = 10) and 5 days' data output*

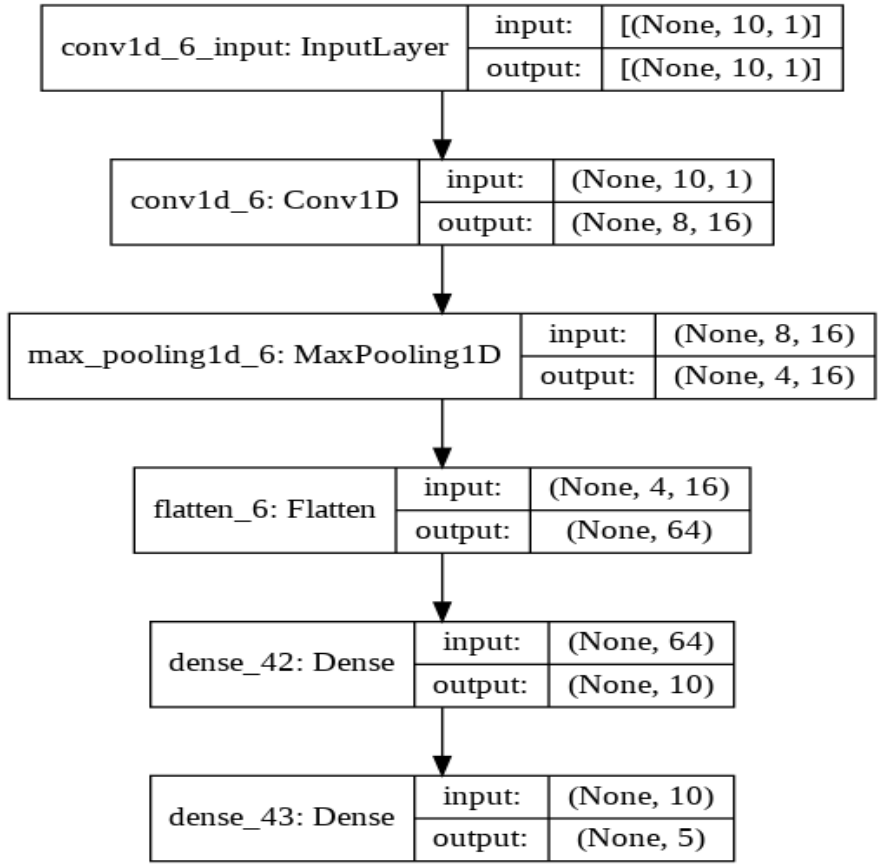

### 4.1.3 Sector-wise results and analysis

The performance of LSTM and CNN networks with 5 inputs and 10 inputs are analyzed in this section. Variation in the RMSE/mean value and the time taken for execution w.r.t the increase or decrease in parameters in both the models are also analyzed. The models are run using the CPU. Day wise RMSE/mean is also plotted and the pattern is analyzed for stocks from different sectors.

### 4.1.4  Metal Sector

**Tata Steel**

*Table 4.1: Tata Steel: LSTM and CNN model performance*

| Model | Parameters | RMSE / Mean % | Exec. Time(sec) |
|---|---|---|---|
| LSTM_UNIV_5 | 142,035 | 5.81 | 82.22 |
| LSTM_UNIV_5 | 182235 | 5.47 | 83.89 |
| LSTM_UNIV_5 | 227,435 | 6.15 | 95.93 |
| LSTM_UNIV_5 | 277,635 | 6.21 | 106.08 |
| LSTM_UNIV_10 | 182235 | 6.68 | 130.12 |
| CNN_UNIV_5 | 233 | 7.35 | 15.71 |
| CNN_UNIV_5 | 289 | 7.20 | 20.93 |
| CNN_UNIV_5 | 345 | 6.83 | 17.53 |
| CNN_UNIV_10 | 769 | 7.90 | 16.08 |

It is evident from the results that the time taken for the execution of CNN is much less compared to that of the LSTM models. There is also a gradual increase in the time taken for execution as the number of parameters is increased in both the LSTM and CNN models. The parameters in the LSTM model are altered by varying the nodes in the LSTM layer whereas in CNN the same is done by varying the no of filters used in the convolutional layer. As far as RMSE/mean is concerned, the change in the parameters does not show any significant difference. But both CNN and LSTM models with the input of N=5 performs better than that with the input of N=10 w.r.t RMSE/mean and execution time

*Figure 4.5: Tata Steel: Day-wise RMSE/Mean Plot(LSTM, N=5)*

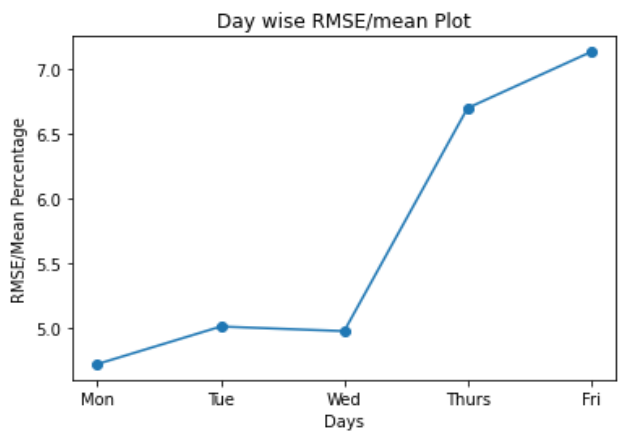

The figure above shows the distribution of day-wise RMS/mean value. From the plot, it is clear that as we move from Monday to Friday there is a gradual increase in the RMSE/ mean value, though Wednesday's value is slightly lesser than that of Tuesday

**JSW Steel**

*Table 4.2: JSW Steel: LSTM and CNN model performance*

| Model | Parameters | RMSE / Mean % | Exec. Time(sec) |
|-------|-----------|---------------|-----------------|
| LSTM_UNIV_5 | 142,035 | 5.27 | 72.53 |
| LSTM_UNIV_5 | 182235 | 5.92 | 88.65 |
| LSTM_UNIV_5 | 227,435 | 6.27 | 96.86 |
| LSTM_UNIV_5 | 277,635 | 5.92 | 115.03 |
| LSTM_UNIV_10 | 182235 | 7.20 | 136.25 |
| CNN_UNIV_5 | 233 | 6.41 | 16.28 |
| CNN_UNIV_5 | 289 | 6.28 | 26.27 |
| CNN_UNIV_5 | 345 | 6.26 | 22.22 |
| CNN_UNIV_10 | 769 | 7.24 | 26.45 |

As observed previously for Tata Steel, the time taken for the execution of CNN is less compared to that of the LSTM models. As the number of parameters is increased, execution time also increases for both the LSTM and CNN models (with an exception of the CNN_UNIV_5 model with 289 parameters) and does not have much impact on RMSE/mean value. Both CNN and LSTM models with the input of N=5 perform better than that with the input of N=10 w.r.t RMSE/mean and execution time.

### 4.1.5  Pharma Sector

**Sun Pharma**

*Table 4.3: Sun Pharma: LSTM and CNN model performance*

| Model | Parameters | RMSE / Mean % | Exec. Time(sec) |
|---|---|---|---|
| LSTM_UNIV_5 | 142,035 | 3.80 | 50.60 |
| LSTM_UNIV_5 | 182235 | 4.20 | 60.46 |
| LSTM_UNIV_5 | 227,435 | 3.98 | 65.10 |
| LSTM_UNIV_5 | 277,635 | 3.78 | 74.90 |
| LSTM_UNIV_10 | 182235 | 3.64 | 90.58 |
| CNN_UNIV_5 | 233 | 4.82 | 13.27 |
| CNN_UNIV_5 | 289 | 4.25 | 13.39 |
| CNN_UNIV_5 | 345 | 4.41 | 11.96 |
| CNN_UNIV_10 | 769 | 3.72 | 13.54 |

Similar to earlier results, the time taken for the execution of CNN is less compared to that of the LSTM models. As the number of parameters is increased, execution time also increases for both the LSTM and CNN models (with an exception of the CNN_UNIV_5 model with 345 parameters) and does not have much impact on RMSE/mean value. Both CNN and LSTM models with the input of N=5 perform better than that with the input of N=10 w.r.t execution time.

The figure (Fig 5.6) shows that there is a gradual increase in RMSE/mean value as we move from Monday to Friday, though Wednesday's RMSE/mean value is slightly lesser than that of Thursday

*Figure 4.6: Sun Pharma: Day wise RMSE/Mean Plot (LSTM, N=5)*

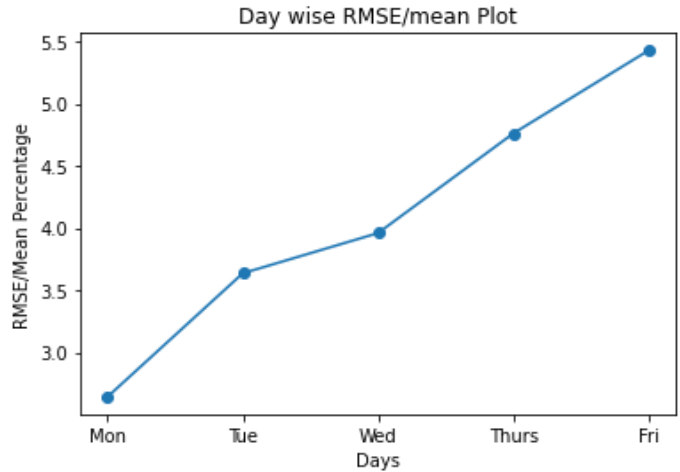

**Divis' Lab**

*Table 4.4: Divis' Lab: LSTM and CNN model performance*

| Model | Parameters | RMSE / Mean % | Exec. Time(sec) |
|---|---|---|---|
| LSTM_UNIV_5 | 142,035 | 3.42 | 75.80 |
| LSTM_UNIV_5 | 182235 | 3.48 | 85.81 |
| LSTM_UNIV_5 | 227,435 | 3.21 | 98.64 |
| LSTM_UNIV_5 | 277,635 | 3.46 | 116.05 |
| LSTM_UNIV_10 | 182235 | 4.42 | 140.45 |
| CNN_UNIV_5 | 233 | 3.67 | 21.33 |
| CNN_UNIV_5 | 289 | 3.72 | 22.55 |
| CNN_UNIV_5 | 345 | 3.82 | 28.93 |
| CNN_UNIV_10 | 769 | 4.41 | 35.09 |

The time taken for the execution of CNN is less compared to that of the LSTM models. As the number of parameters is increased, execution time also increases for both the LSTM and CNN models and does not have much impact on RMSE/mean value. Both CNN and LSTM models with the input of N=5 perform better than that with the input of N=10 w.r.t RMSE/mean value and execution time.

### 4.1.6   **IT Sector**

**Infosys**

*Table 4.5: Infosys: LSTM and CNN model performance*

| Model | Parameters | RMSE / Mean % | Exec. Time(sec) |
|-------|-----------|---------------|-----------------|
| LSTM_UNIV_5 | 142,035 | 3.52 | 52.92 |
| LSTM_UNIV_5 | 182235 | 3.40 | 63.47 |
| LSTM_UNIV_5 | 227,435 | 4.29 | 67.79 |
| LSTM_UNIV_5 | 277,635 | 3.88 | 76.67 |
| LSTM_UNIV_10 | 182235 | 6.24 | 95.13 |
| CNN_UNIV_5 | 233 | 4.28 | 13.27 |
| CNN_UNIV_5 | 289 | 3.79 | 14.84 |
| CNN_UNIV_5 | 345 | 3.77 | 13.75 |
| CNN_UNIV_10 | 769 | 5.24 | 12.56 |

The time taken for the execution of CNN is less compared to that of the LSTM models. As the number of parameters is increased, execution time also increases for both the LSTM and CNN models (with the exception of the CNN_UNIV_5 model with 345 parameters and CNN_UNIV_10 model) and does not have much impact on RMSE/mean value. Both CNN and LSTM models with the input of N=5 perform better than that with the input of N=10 w.r.t RMSE/mean value and RMSE/mean value. Interestingly CNN_UNIV_10 has taken the least execution time compared to all other CNN_UNIV_5 models

*Figure 4.7: Infosys: Day-wise RMSE/Mean Plot (LSTM, N=5)*

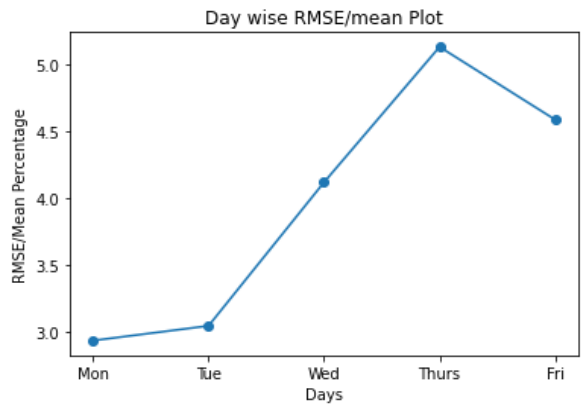

Day wise RMSE/Mean plot shows that there is a gradual increase in RMSE/mean value as we move from Monday to Friday, though Tuesday's and Friday's RMSE/mean value is slightly lesser than that of their previous days.

**TCS**

*Table 4.6: TCS: LSTM and CNN model performance*

| Model | Parameters | RMSE / Mean % | Exec. Time(sec) |
|---|---|---|---|
| LSTM_UNIV_5 | 142,035 | 3.33 | 68.35 |
| LSTM_UNIV_5 | 182235 | 3.47 | 83.43 |
| LSTM_UNIV_5 | 227,435 | 3.44 | 90.31 |
| LSTM_UNIV_5 | 277,635 | 3.12 | 105.67 |
| LSTM_UNIV_10 | 182235 | 3.67 | 127.68 |
| CNN_UNIV_5 | 233 | 3.71 | 16.30 |
| CNN_UNIV_5 | 289 | 3.34 | 21.49 |
| CNN_UNIV_5 | 345 | 3.36 | 25.94 |
| CNN_UNIV_10 | 769 | 3.88 | 26.70 |

Consistent with some of the previous results, the time taken for the execution of CNN is less compared to that of the LSTM models. As the number of parameters is increased, execution time also increases for both the LSTM and CNN models and does not have much impact on RMSE/mean value. Both CNN and LSTM models with the input of N=5 perform better than that with the input of N=10 w.r.t RMSE/mean value and execution time.

### 4.1.7 Banking Sector

**HDFC Bank**

*Table 4.7: HDFC Bank: LSTM and CNN model performance*

| Model | Parameters | RMSE / Mean % | Exec. Time(sec) |
|---|---|---|---|
| LSTM_UNIV_5 | 142,035 | 3.78 | 57.62 |
| LSTM_UNIV_5 | 182235 | 3.76 | 68.42 |
| LSTM_UNIV_5 | 227,435 | 3.92 | 94.08 |
| LSTM_UNIV_5 | 277,635 | 3.92 | 104.84 |
| LSTM_UNIV_10 | 182235 | 4.31 | 111.11 |
| CNN_UNIV_5 | 233 | 3.91 | 14.22 |
| CNN_UNIV_5 | 289 | 3.97 | 15.32 |
| CNN_UNIV_5 | 345 | 4.26 | 15.06 |
| CNN_UNIV_10 | 769 | 4.83 | 21 |

The time taken for the execution of CNN is less compared to that of the LSTM models. As the number of parameters is increased, execution time also increases for both the LSTM and CNN models (with the exception of the CNN_UNIV_5 model with 345) and does not have much impact on RMSE/mean value. Both CNN and LSTM models with the input of N=5 perform better than that with the input of N=10 w.r.t RMSE/mean value and RMSE/mean value.

*Figure 4.8: HDFC Bank: Day-wise RMSE/Mean Plot (LSTM, N=5)*

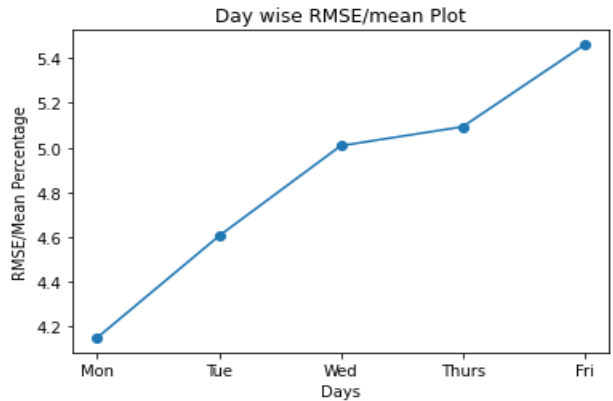

The plot (Fig 5.8) shows that the RMSE/Mean value is increasing as we move from Monday to Friday, though Thursday is an exception

**ICICI Bank**

*Table 4.8: ICICI Bank: LSTM and CNN model performance*

| Model | Parameters | RMSE / Mean % | Exec. Time(sec) |
|---|---|---|---|
| LSTM_UNIV_5 | 142,035 | 4.80 | 85.37 |
| LSTM_UNIV_5 | 182235 | 4.85 | 100.76 |
| LSTM_UNIV_5 | 227,435 | 4.59 | 114.49 |
| LSTM_UNIV_5 | 277,635 | 4.91 | 118.30 |
| LSTM_UNIV_10 | 182235 | 5.57 | 114.86 |
| CNN_UNIV_5 | 233 | 5.00 | 21.73 |
| CNN_UNIV_5 | 289 | 5.07 | 22.38 |
| CNN_UNIV_5 | 345 | 5.08 | 24.56 |
| CNN_UNIV_10 | 769 | 5.10 | 25.50 |

The time taken for the execution of CNN is less compared to that of the LSTM models. As the number of parameters is increased, execution time also increases for both the LSTM and CNN models (with the exception of the CNN_UNIV_10 model) and does not have much impact on RMSE/mean value. CNN models with the input of N=5 perform better than that with the input of N=10 w.r.t RMSE/mean value and execution time. Noticeably, LSTM_UNIV_10 has taken the least less execution time compared to one of the LSTM_UNIV_5 models

### 4.1.8 Auto Sector

**Maruti Suzuki**

*Table 4.9: Maruti Suzuki: LSTM and CNN model performance*

| Model | Parameters | RMSE / Mean % | Exec. Time(sec) |
|---|---|---|---|
| LSTM_UNIV_5 | 142,035 | 3.88 | 49.94 |
| LSTM_UNIV_5 | 182235 | 4.73 | 59.33 |
| LSTM_UNIV_5 | 227,435 | 4.11 | 63.85 |
| LSTM_UNIV_5 | 277,635 | 3.99 | 73.00 |
| LSTM_UNIV_10 | 182235 | 5.42 | 90.29 |
| CNN_UNIV_5 | 233 | 4.47 | 11.30 |
| CNN_UNIV_5 | 289 | 4.25 | 13.05 |
| CNN_UNIV_5 | 345 | 4.28 | 12.03 |
| CNN_UNIV_10 | 769 | 4.81 | 13.43 |

As observed before, the time taken for the execution of CNN is less compared to that of the LSTM models. As the number of parameters is increased, execution time also increases for both the LSTM and CNN models (with the exception of the CNN_UNIV_5 model with 345 parameters) and does not have much impact on RMSE/mean value. Both CNN and LSTM models with the input of N=5 perform better than that with the input of N=10 w.r.t RMSE/mean value and execution time.

Day-wise RMSE/Mean plot (Fig 33) shows that there is a gradual increase in RMSE/mean value as we move from Monday to Friday, though Wednesday's RMSE/mean value is slightly lesser than that of the previous day.

*Figure 4.9: Maruti Suzuki: Day wise RMSE/Mean Plot (LSTM, N=5)*

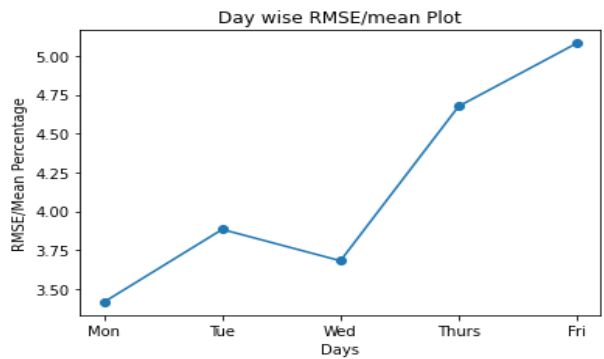

**Mahindra & Mahindra**

*Table 4.10: Mahindra & Mahindra: LSTM and CNN model performance*

| Model | Parameters | RMSE / Mean % | Exec. Time(sec) |
|---|---|---|---|
| LSTM_UNIV_5 | 142,035 | 4.65 | 69.48 |
| LSTM_UNIV_5 | 182235 | 4.54 | 86.05 |
| LSTM_UNIV_5 | 227,435 | 4.78 | 92.77 |
| LSTM_UNIV_5 | 277,635 | 4.88 | 108.66 |
| LSTM_UNIV_10 | 182235 | 4.83 | 128.84 |
| CNN_UNIV_5 | 233 | 5.11 | 21.55 |
| CNN_UNIV_5 | 289 | 5.04 | 26.32 |
| CNN_UNIV_5 | 345 | 5.18 | 21.94 |
| CNN_UNIV_10 | 769 | 4.98 | 22.03 |

The time taken for the execution of CNN is less compared to that of the LSTM models. As the number of parameters is increased, execution time also increases for both the LSTM and CNN models (with the exception of the CNN_UNIV_5 model with 289 parameters) and does not have much impact on RMSE/mean value. LSTM models with the input of N=5 perform better than that with the input of N=10 w.r.t RMSE/mean value and execution time.

By studying 10 stocks from 5 sectors and running LSTM and CNN models on them, it is clear that CNN models always take less execution time compared to LSTM models but with a trade off in performance. Overall it's LSTM that s performing better than LSTM, even while there is change in parameters or inputs. Also, LSTM and CNN models with an input of the past one week's data performs better than the past two week's data

# Chapter 6

## Portfolio Optimization

Chapters 3 to 6 discussed the various statistical, econometric, machine learning, and deep learning models to predict stock prices of selected stocks from five sectors. This chapter discusses a systematic approach for building robust portfolios of stocks -minimum risk portfolio, optimal risk portfolio- from those five sectors. Five stocks from each sector would be considered for portfolio building. Historical stock prices of five years that are from Jan 1, 2016, to Aug 31, 2020, would be used for building the portfolios. The performances of the portfolios are evaluated based on their returns after a period of eight months. Several other attributes of the portfolios such as the risk, weights assigned to different stocks, and the correlation among the constituent stocks are also studied.

**Steps followed in portfolio design**

I. **Data acquisition of 5 stocks from each of the chosen five sectors**

As discussed in Chapter 2, using Yahoo Finance API, stock prices of five years i.e. from Jan 1, 2016, to Aug 31, 2020, are fetched for building the portfolio and from Jan 1, 2021, to Aug 31, 2021, for backtesting. Since the focus is on univariate analysis, 'close price' is the variable of interest.

II. **Computation of return and volatility**

The daily return and log return values of each stock of the sector is calculated. The daily return values are the percentage changes in the daily close values over successive days, while the log return values are the logarithms of the percentage changes in the daily close values. Using the daily return values, the daily volatility and the annual volatility of the five stocks of each sector are computed. The daily volatility is defined as the standard deviation of the daily return values. The daily volatility, on multiplication by a factor of the square root of 250, yields the value of the annual volatility. The annual volatility value of a stock quantifies the risk associated with stock from the point of view of an investor, as it indicates the amount of variability in its price. The daily return values are also aggregated into annual return values for each stock for every sector.

### III.    Computation of covariance and correlation matrices

Once the volatilities and return of the stocks are computed, the covariance and the correlation matrices for the five stocks in each sector are calculated. These matrices help us in understanding the strength of association between a pair of stock prices in a given sector. Any pair exhibiting a high value of correlation coefficient indicates a strong association between them. A good portfolio aims to minimize the risk while optimizing the return. Risk minimization of a portfolio requires identifying stocks that have low correlation among themselves so that a higher diversity can be achieved. Hence, computation and analysis of the covariance and correlation matrices of the stocks are of importance.

### IV.    Computation of the expected return and risk of portfolios

At this step, we proceed towards a deeper analysis of the historical prices of the five stocks in each of the five sectors. First, for each sector, we construct a portfolio using the five stocks, with each stock carrying equal weight. Since there are five stocks in a sector (i.e., in a portfolio), each stock is assigned a weight of 0.2. Based on the training dataset and using an equal-weight portfolio, we compute the yearly return and risk (i.e., volatility) of each portfolio.

The yearly return and the yearly volatility of the equal-weight portfolio of each sector are computed using the training dataset. For this purpose, the mean of the yearly return values is derived using the resample function in Python with a parameter 'Y'. Yearly volatility values of the stocks in the equal-weight portfolio are derived by multiplying the daily volatility values by the square root of 250, assuming that there are, on average, 250 working days in a year for a stock exchange. The equal-weight portfolio of a sector gives us an idea about the overall profitability and risk associated with each sector over the training period. However, for future investments, their usefulness is very limited. Every stock in a portfolio does not contribute equally to its return and the risk. Hence, we proceed with computations of minimum risk and optimal risk portfolios in the next steps.

V.    **Building the minimum risk portfolio**

We build the minimum risk portfolio for each sector using the records in its training dataset. The minimum risk portfolio is characterized by its minimum variance. The variance of a portfolio is a metric computed using the variances of each stock in the portfolio as well as the covariance between each pair of stocks in the portfolio.

For finding the minimum risk portfolio, we first plot the efficient frontier for each portfolio. For a given portfolio of stocks, the efficient frontier is the contour with returns plotted along the y-axis and the volatility (i.e., risk) on the x-axis. The points of an efficient frontier denote the points with the maximum return for a given value of volatility or the minimum value of volatility for a given value of the return. Since, for an efficient frontier, the volatility is plotted along the x-axis, the minimum risk portfolio is identified by the leftmost point lying on the efficient frontier. For plotting the contour of the efficient frontier, we randomly assign the weights to the five stocks in a portfolio in a loop and produce 10,000 points through iteration, each point representing a portfolio. The minimum risk portfolio is identified by detecting the leftmost point on the efficient frontier.

## VI. Computing the optimal risk portfolio

The investors in the stock markets are usually not interested in the minimum risk portfolios as the return values are usually low. In most cases, the investors are ready to incur some amount of risk if the associated return values are high. To compute the optimum risk portfolio, we use the metric Sharpe Ratio of a portfolio. The Sharpe Ratio of a portfolio is given below.

$$Sharpe\ Ratio = (Rc - Rf) / \sigma c$$

$Rc$, $Rf$, and $\sigma c$ denote the return of the current portfolio, the risk-free portfolio, and the standard deviation of the current portfolio, respectively. Here, the risk-free portfolio is a portfolio with a volatility value of 1%. The optimum-risk portfolio is the one that maximizes the Sharpe Ratio for a set of stocks. This portfolio makes an optimization between the return and the risk of a portfolio. It yields a substantially higher return than the minimum risk portfolio, with a very nominal increase in the risk, and hence, maximizing the value of the Sharpe ratio.

## 4.2 Sector-wise results

In this section, the results of the five portfolios and analysis is given. As mentioned earlier, the chosen five sectors are (i) metal, (ii) banking, (iii) information technology (IT), (iv) banking, and (v) auto

### 4.2.1 Metal Sector

Table 6.1 depicts the annual return and the annual risk (i.e. volatility) for the stocks of the metal sector over the training period, i.e., from 1 January 2016 to values 27 December 2020. Coal India is found to exhibit the lowest annual return and the lowest annualized risk. While Adani Enterprises yields the highest annual return and the highest risk.

*Table 0.1: Return and risk of metal sector stocks*

| Stocks | Annual Return (%) | Annual Risk (%) |
|---|---|---|
| Tata Steel | 22.20 | 38.07 |
| Hindalco | 16.49 | 41.68 |
| JSW Steel | 27.72 | 36.25 |
| Adani Enterprises | 88.62 | 51.27 |
| Coal India | -17.22 | 29.40 |

Table 6.2 presents the allocation of weights to different stocks of the metal sector using two portfolio design approaches – (i) minimum risk portfolio and (ii) optimum risk portfolio. The sum of weights for each of the three cases is 1. The stock which receives the highest allocation of weight as per the minimum risk and optimum risks are Coal India and Adani enterprises respectively.

*Table 0.2: The portfolios of the metal sector stocks*

| Stocks | Min Risk Portfolio | Opt. Risk Portfolio |
|---|---|---|
| Tata Steel | 0.05 | 0.08 |
| Hindalco | 0.01 | 0.05 |
| JSW Steel | 0.30 | 0.01 |
| Adani Enterprises | 0.06 | 0.84 |
| Coal India | 0.56 | 0.01 |

Table 6.3 shows the risk and the return values associated with the two portfolios of the metal sector stocks. These values are computed using the prices of the stocks over the training period. It is found that between the two portfolios, the optimum risk portfolio yielded the highest values for the return and risk.

*Table 0.3: The return and the risk values of the metal sector portfolios*

| Metric | Min Risk Portfolio | Opt. Risk Portfolio |
|---|---|---|
| Portfolio Return | 5.78% | 77.36% |
| Portfolio Risk | 25.90% | 45.74% |

Table 6.4 shows the return for an investor who followed the optimum risk portfolio approach and invested a total amount of INR 100000 on 1 January 2021. Note that the total amount here is just an example. The overall return percent will not be affected by the amount invested. As per Table 6.4, an investor who followed the optimum risk portfolio approach would invest the respective amount in the stocks based on the proportion indicated by the portfolio design. These values are noted under the column "Amount Invested" on 1 January 2021. The market values of the stocks are noted in the column "Price/Stock" on 1 January 2021. Using these values, the number of shares purchased by the investor for each stock are computed and are listed under the column "No. of Stocks". After six months, the actual price of each stock is noted and listed in the column "Price/Stock" on 1 July 2021. For a given stock, its actual price on 1 July 2021 is multiplied with the corresponding no. of shares to compute the actual value of the stocks. The actual values of all five stocks are summed up to find the total value of the stocks on 1 July 2021. The return for the eight months' period under the optimum risk portfolio is found to be 188.95%. Figure 6.4 exhibits the efficient frontier of the portfolios of the metal sector stocks.

*Table 0.4: The actual return of the optimum portfolio of the metal sector*

| Stock | Date: January 1, 2021 | | | Date: August 31, 2021 | | Return |
|---|---|---|---|---|---|---|
| | Price/ Stock | Amount Invested | Volume of Stocks | Price/ Stock | Actual Value of Stocks | |
| Tata Steel | 643 | 8481.35 | 13.19 | 1384 | 18255.35 | |
| Hindalco | 238 | 5266.11 | 22.12 | 438 | 9691.43 | |
| JSW Steel | 390 | 255.43 | 0.65 | 678 | 444.06 | 188.95% |
| Adani Enterprises | 491 | 84421.28 | 171.93 | 1506 | 258937.79 | |
| Coal India | 135 | 1575.80 | 11.67 | 139 | 1622.49 | |

The figure below (Fig 6.1) exhibits the efficient frontier of the portfolios of the metal sector stocks

*Figure 0.1: The minimum risk portfolio (the red star) and the optimum risk portfolio (the green star) for the metal sector on historical stock prices from 1 January 2016 to 27 December 2020 (The risk is plotted along the x-axis and the return along the y-axis)*

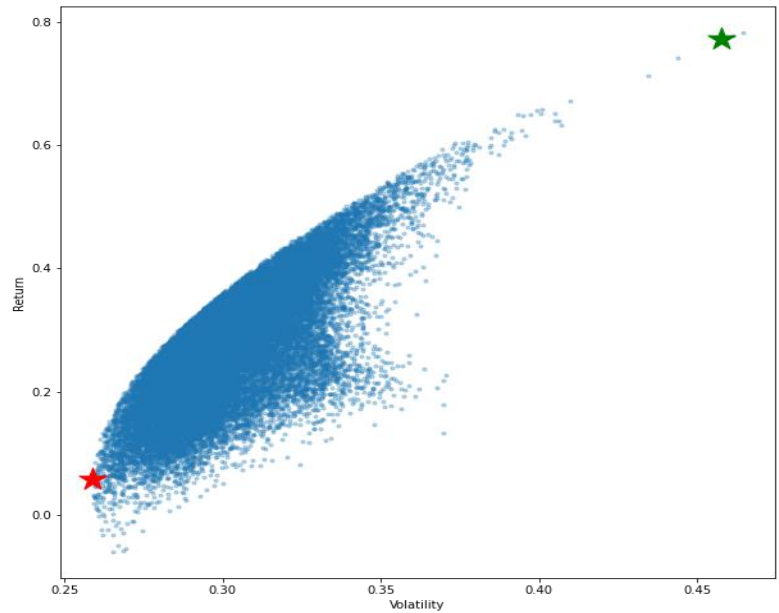

The figure (Fig 6.2) depicts the returns of all the three portfolios during the training period. The graph shows that both the optimal risk portfolio and minimum variance portfolio obtained returns higher than the equal weight portfolio, as expected. The performance of the portfolios can be better compared using Sharpe ratio. Table 6.5 shows that the Sharpe ratio of Optimal portfolio is much higher than equal weight and minimum risk portfolios. In the figure (Fig 6.3) backtesting of the portfolios built is depicted. The return obtained from the portfolio, if the portfolio was sold on each day after January 1, 2021, till Aug 31 2021 is calculated for all three portfolios. The equal-weight portfolio is being taken as the benchmark or it is the portfolio in which a naive investor would invest. The graph shows that both the optimal risk portfolio and minimum variance portfolio obtained returns higher than the equal weight portfolio, as expected. From table 6.6, it is clear that Sharpe ratio of optimal risk portfolio is higher than the other two as in the in-sample results.

*Figure 0.2: Comparison of Return% of Equal Weight, minimum variance, and Optimal Risk portfolio for in-sample*

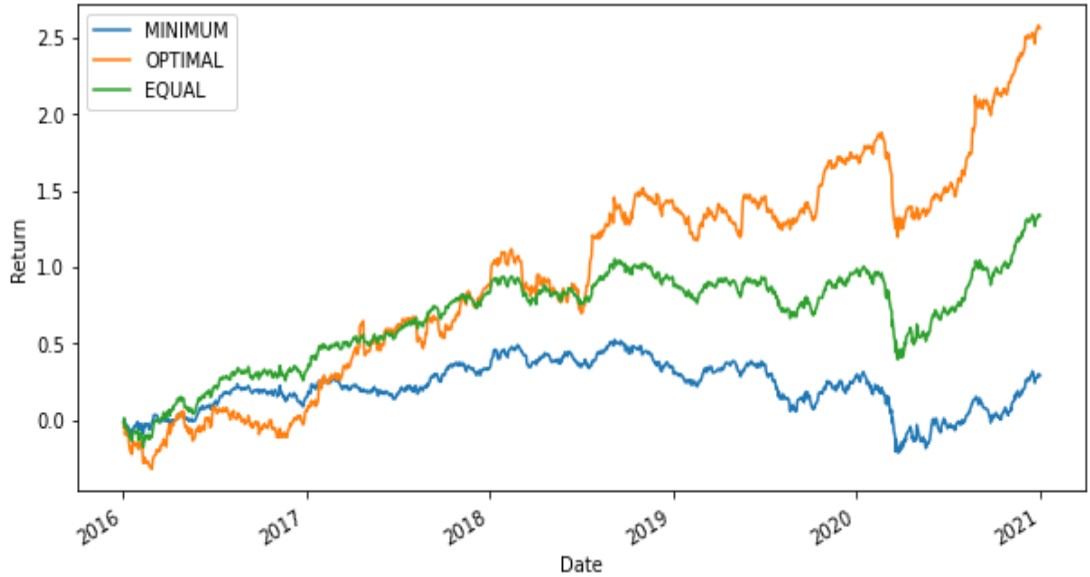

*Figure 0.3: Comparison of Return% of Equal Weight, minimum variance, and Optimal Risk portfolio for out of sample*

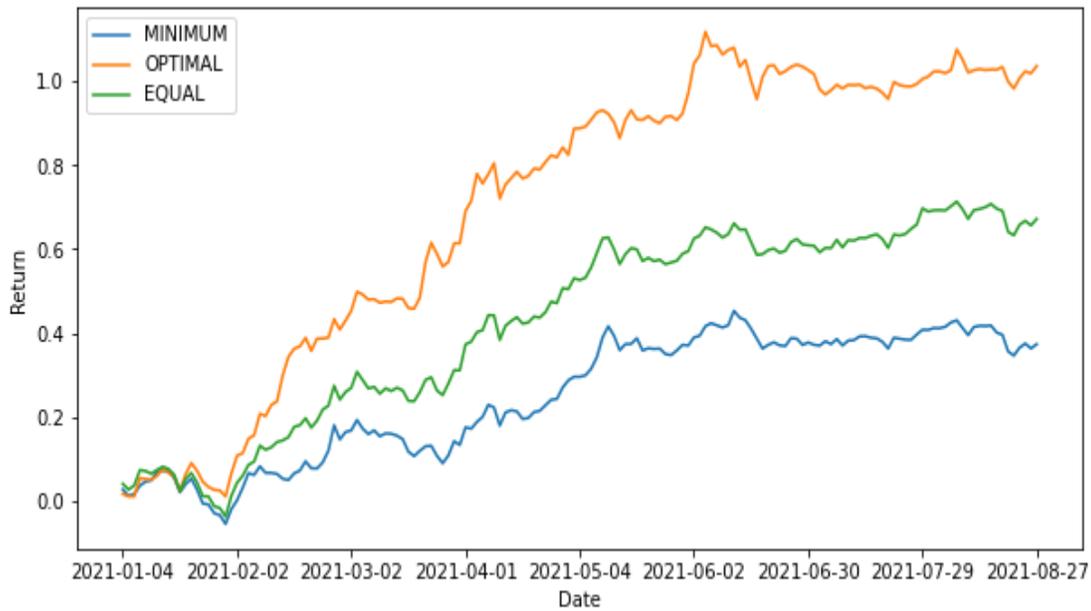

*Table 0.5: In-sample results*

| Portfolio | Return | Stdev | Sharpe Ratio |
|-----------|--------|-------|--------------|
| **MINIMUM** | 3.75 | 0.25 | 0.23 |
| **OPTIMAL** | 33.03 | 0.43 | 1.19 |
| **EQUAL** | 17.23 | 0.29 | 0.92 |

*Table 0.6: Out of sample results*

| Portfolio | Return | Stdev | Sharpe Ratio |
|-----------|--------|-------|--------------|
| **MINIMUM** | 33.00 | 0.27 | 1.87 |
| **OPTIMAL** | 106.56 | 0.43 | 3.92 |
| **EQUAL** | 66.25 | 0.31 | 3.34 |

If the return the portfolio would yield could be predicted accurately for a future date, it would help the investor or buyer of a portfolio to take intuitive sell/buy decisions on the portfolio. The stock price value for the next day for all the stocks in the portfolio is calculated using an LSTM model and the portfolio returns are calculated using the predicted price. A comparison of optimal risk portfolio return % obtained using actual stock price and predicted stock price is calculated and plotted. An LSTM model is designed for predicting future stock prices. The model uses the daily close price of the stock of the past 50 days as the input. The prediction is done for a single day. The graph (Fig 6.4) shows that both the lines closely follow each other which shows the precision of prediction of stock price and its practical application in predicting future portfolio value.

*Figure 0.4: Comparison of Return% of Optimal Risk portfolio on actual stock prices and predicted stock prices*

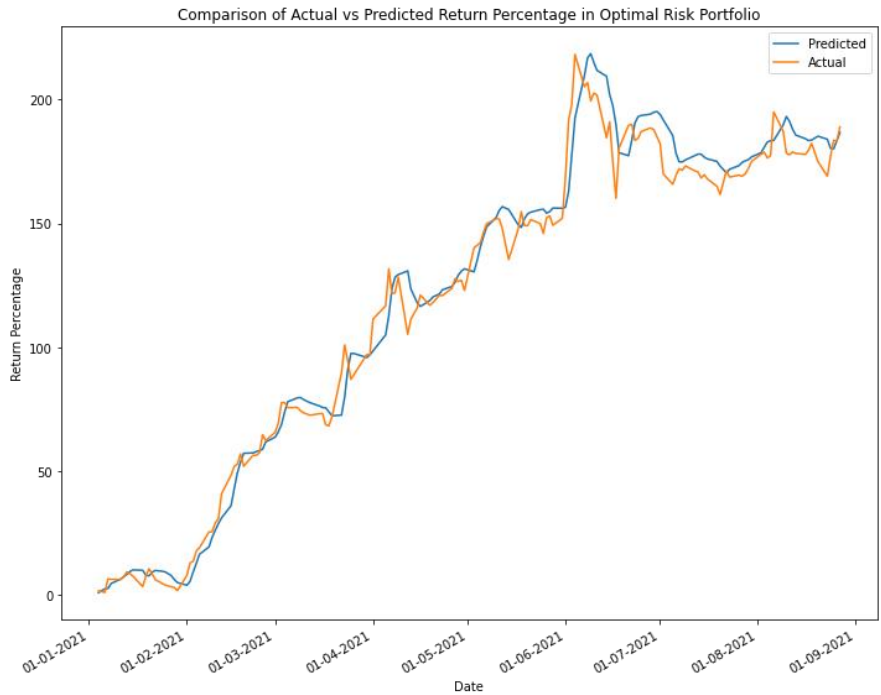

### 4.2.2   Pharma Sector

*Table 0.7: Return and risk of pharma sector stocks*

| Stocks | Annual Return (%) | Annual Risk (%) |
|---|---|---|
| Sun Pharma | 0.86 | 32.94 |
| Divi's Lab | 51.94 | 36.08 |
| Dr. Reddy's Laboratories | 19.55 | 28.94 |
| Cipla | 13.96 | 27.98 |
| Lupin | -6.67 | 31.46 |

Table 6.7 depicts the annual return and the annual risk (i.e. volatility) for the stocks of the pharma sector over the training period, i.e., from 1 January 2016 to values 27 December 2020. Lupin is found to exhibit the lowest annual return and Cipla the lowest annualized risk. While Divi's Lab yields the highest annual return and the highest risk.

Table 6.8 presents the allocation of weights to different stocks of the pharma sector using two portfolio design approaches – (i) minimum risk portfolio and (ii) optimum risk portfolio. The sum of weights for each of the three cases is 1. The stocks which receive the highest allocation of weight as per the minimum risk and optimum risks are Cipla and Divi's Lab respectively.

*Table 0.8: The portfolios of the pharma sector stocks*

| Stocks | Min Risk Portfolio | Opt. Risk Portfolio |
|---|---|---|
| Sun Pharma | 0.12 | 0.01 |
| Divi's Lab | 0.14 | 0.73 |
| Dr. Reddy's Lab | 0.29 | 0.16 |
| Cipla | 0.30 | 0.07 |
| Lupin | 0.13 | 0.01 |

Table 6.9 shows the risk and the return values associated with the two portfolios of the pharma sector stocks. These values are computed using the prices of the stocks over the training period. It is found that between the two portfolios, the optimum risk portfolio yielded the highest values for the return and risk.

*Table 0.9: The return and the risk values of the IT sector portfolios*

| Metric | Min Risk Portfolio | Opt. Risk Portfolio |
|--------|--------------------|--------------------|
| Portfolio Return | 16.90% | 42.55% |
| Portfolio Risk | 22.18% | 29.81% |

Table 6.10 shows the return for an investor who followed the optimum risk portfolio approach and invested a total amount of INR 100000 on 1 January 2021. The return for the eight months' period under the optimum risk portfolio is found to be 19.45%. Figure 6.8 exhibits the efficient frontier of the portfolios of the Pharma sector stocks.

*Table 0.10: The actual return of the optimum portfolio of the Pharma sector*

| Stock | Date: January 1, 2021 | | | Date: August 31, 2021 | | Return |
|-------|-----------|--------|-----------|------------|-----------|--------|
| | Price/ Stock | Amount Invested | Volume of Stocks | Price/Stock | Actual Value of Stocks | |
| Sun Pharma | 596 | 786.22 | 1.31 | 772 | 1018.40 | 19.45% |
| Divi's Lab | 3849 | 73946.66 | 19.21 | 4911 | 94349.72 | |
| Dr. Reddy's Laboratories | 5241 | 16262.84 | 3.10 | 4601 | 14276.91 | |
| Cipla | 827 | 7573.12 | 9.15 | 924 | 8461.38 | |
| Lupin | 1001 | 1431.15 | 1.42 | 940 | 1343.93 | |

The figure below (Fig 6.5) exhibits the efficient frontier of the portfolios of the pharma sector stocks

*Figure 0.5: The minimum risk portfolio (the red star) and the optimum risk portfolio (the green star) for the pharma sector on historical stock prices from 1 January 2016 to 27 December 2020 (The risk is plotted along the x-axis and the return along the y-axis)*

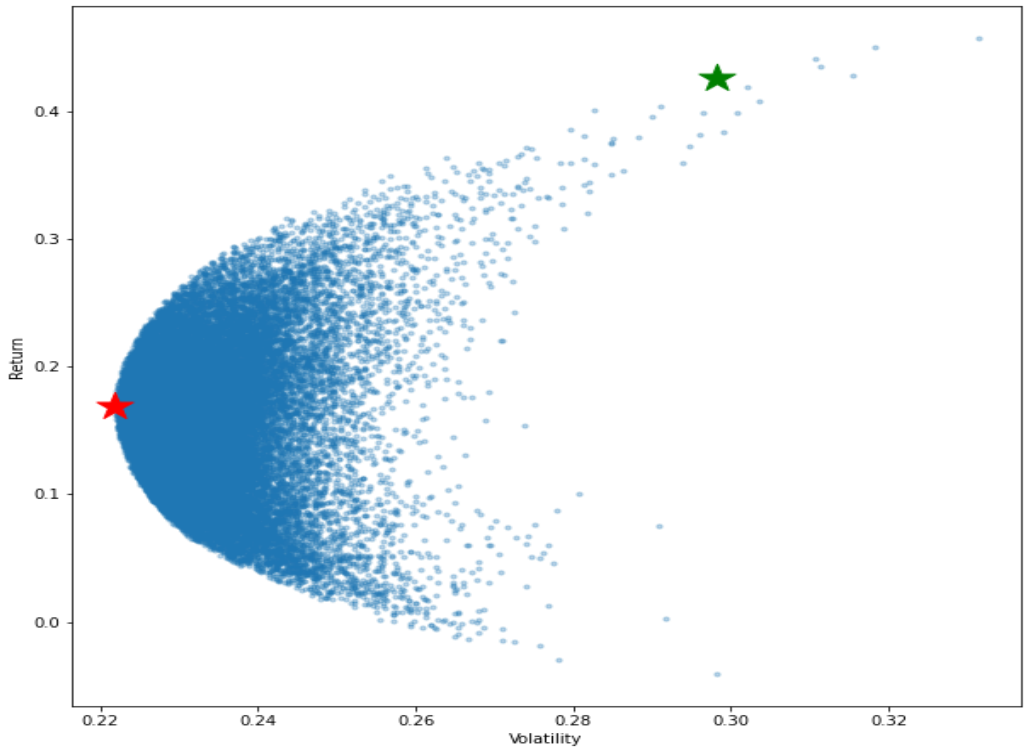

The figure (Fig 6.6) depicts the returns of all the three portfolios during the training period. The graph shows that the optimal risk portfolio obtained returns higher than the minimum variance portfolio and the equal weight portfolio, as expected. Table 6.11 also shows that the Sharpe ratio of the optimal risk portfolio is higher than the other two. In the figure (Fig 6.7) backtesting of the portfolios built is depicted. The graph shows that the optimal risk portfolio did not perform not even as well as the equal-weight portfolio which can be validated by comparing the Sharpe ratio of the portfolios from Table 6.12. This points towards one of the drawbacks of Optimal Risk Portfolio i.e. there is overfitting with the training data and hence the test performance is not in line with the training results.

*Figure 0.6: Comparison of Return% of Equal Weight, minimum variance, and Optimal Risk portfolio for in-sample*

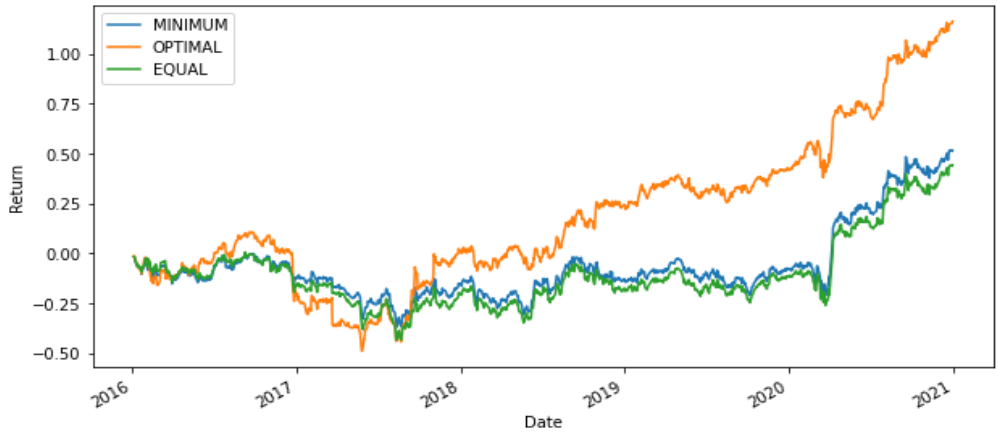

*Figure 0.7: Comparison of Return% of Equal Weight, minimum variance, and Optimal Risk portfolio for out of sample*

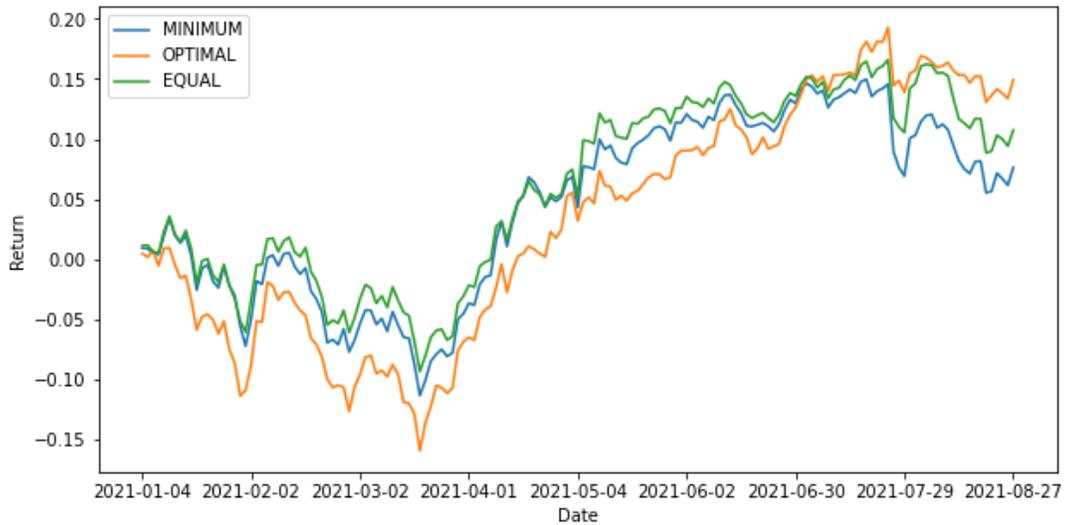

*Table 0.11: In sample results*

| Portfolio | Return | Stdev | Sharpe Ratio |
|-----------|--------|-------|--------------|
| **MINIMUM** | 6.63 | 0.22 | 0.47 |
| **OPTIMAL** | 14.98 | 0.27 | 0.86 |
| **EQUAL** | 5.69 | 0.22 | 0.39 |

*Table 0.12: Out Sample Results*

| Portfolio | Return | Stdev | Sharpe Ratio |
|-----------|--------|-------|--------------|
| **MINIMUM** | 7.52 | 0.20 | 0.58 |
| **OPTIMAL** | 14.71 | 0.20 | 1.16 |
| **EQUAL** | 10.58 | 0.20 | 0.828 |

The graph (Fig 6.8) shows that both the line (portfolio return percentage calculated on actual and predicted stock price) closely follow each other which shows the precision of prediction of stock price

*Figure 0.8: Comparison of Return% of Optimal Risk portfolio on actual stock prices and predicted stock prices for pharma sector*

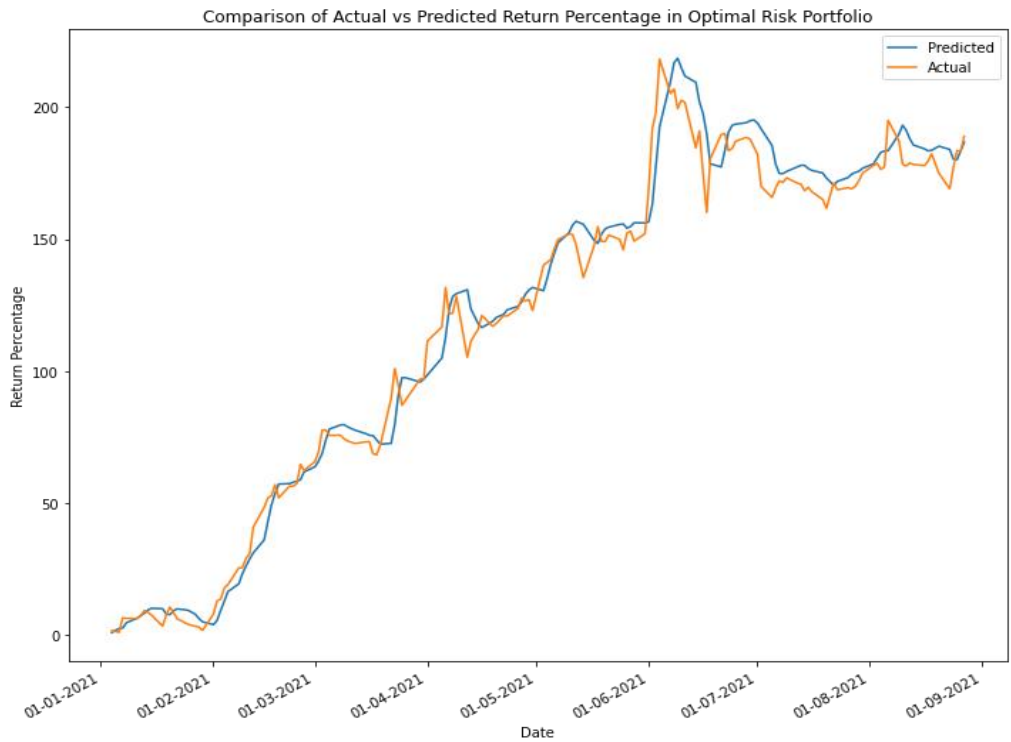

### 4.2.3  IT Sector

Table 6.13 depicts the annual return and the annual risk (i.e., volatility) for the stocks of

the IT sector over the training period. Tech Mahindra is found to exhibit the lowest annual

return and Wipro the lowest annualized risk. Infosys yields the highest annual return and

Tech Mahindra the highest risk.

*Table 0.13: Return and risk of IT sector stocks*

| Stocks | Annual Return (%) | Annual Risk (%) |
|---|---|---|
| Infosys | 28.07 | 28.55 |
| TCS | 25.24 | 25.83 |
| Tech Mahindra | 19.91 | 31.03 |
| Wipro | 23.48 | 25.47 |
| HCL Technologies | 25.05 | 27.55 |

*Table 0.14: The portfolios of the IT sector stocks*

| Stocks | Min Risk Portfolio | Opt. Risk Portfolio |
|---|---|---|
| Infosys | 0.13 | 0.25 |
| TCS | 0.25 | 0.28 |
| Tech Mahindra | 0.11 | 0.01 |
| Wipro | 0.30 | 0.07 |
| HCL Technologies | 0.13 | 0.01 |

Table 6.14 presents the allocation of weights to different stocks of the IT sector using two portfolio design approaches – (i) minimum risk portfolio and (ii) optimum risk portfolio. The stock which receives the highest allocation of weight as per the minimum risk and optimum risks are Wipro and TCS respectively.

Table 6.15 shows the risk and the return values associated with the two portfolios of the IT sector stocks. These values are computed using the prices of the stocks over the training period. It is found that between the two portfolios, the optimum risk portfolio yielded the highest values for the return and risk.

*Table 0.15: The return and the risk values of the IT sector portfolios*

| Metric | Min Risk Portfolio | Opt. Risk Portfolio |
|---|---|---|
| Portfolio Return | 24.38% | 25.38% |
| Portfolio Risk | 20.80% | 21.08% |

Table 6.16 shows the return for an investor who followed the optimum risk portfolio approach and invested a total amount of INR 100000 on 1 January 2021. The return for the eight months' period under the optimum risk portfolio is found to be 39.14%. Figure 6.12 exhibits the efficient frontier of the portfolios of the IT sector stocks.

*Table 0.16: The actual return of the optimum portfolio of the IT sector*

| Stock | Date: January 1, 2021 | | | Date: August 31, 2021 | | Return |
|---|---|---|---|---|---|---|
| | Price/ Stock | Amount Invested | Volume of Stocks | Price/ Stock | Actual Value of Stocks | |
| Infosys | 1260 | 25027.52 | 19.86 | 1709 | 33946.06 | 39.14% |
| TCS | 2928 | 28157.27 | 9.61 | 3720 | 35773.58 | |
| Tech Mahindra | 978 | 392.89 | 0.40 | 1445 | 580.51 | |
| Wipro | 388 | 29131.70 | 75.08 | 635 | 47676.88 | |
| HCL Technologies | 950 | 17290.61 | 18.20 | 1163 | 21167.35 | |

The figure below (Fig 6.9) exhibits the efficient frontier of the portfolios of the IT sector stocks

*Figure 0.9: The minimum risk portfolio (the red star) and the optimum risk portfolio (the green star) for the IT sector on historical stock prices from 1 January 2016 to 27 December 2020 (The risk is plotted along the x-axis and the return along the y-axis)*

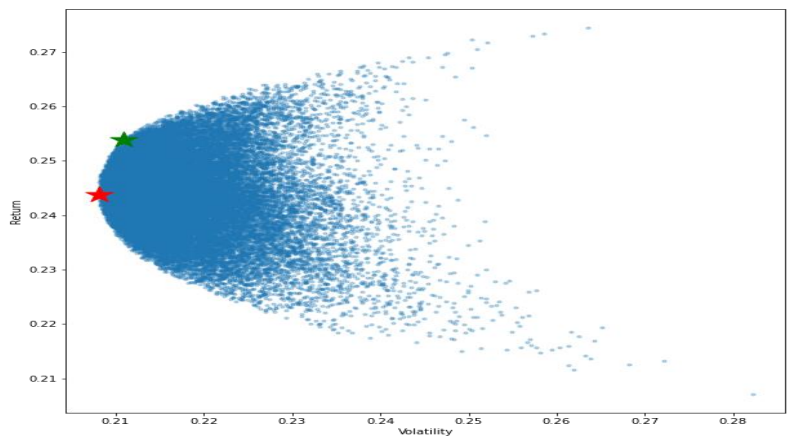

Figure 6.9: The minimum risk portfolio (the red star) and the optimum risk portfolio (the green star) for the IT sector on historical stock prices from 1 January 2016 to 27 December 2020 (The risk is plotted along the x-axis and the return along the y-axis)

The figure (Fig 6.10) depicts the returns of all the three portfolios during the training period. The graph shows that the optimal risk portfolio obtained returns almost equal to slightly higher than the minimum variance portfolio and the equal weight portfolio. From Table 6.17 by comparing the Sharpe ratio of portfolios, the performance of optimal portfolio does not show much variation from that of the other two portfolios.  The high correlation of the stocks in the IT sector lead to the anomalous results. In the figure (Fig 6.11) backtesting of the portfolios built is depicted. The graph shows that the optimal risk portfolio performed better than the equal weight portfolio but not as good as the minimum variance portfolio, which can be validated by comparing the Sharpe ratio of the portfolios from table 6.18. The overfitted train data and the correlation among the stocks in the portfolio caused the failure of Optimal Risk Portfolio

*Figure 0.10: Comparison of Return% of Equal Weight, minimum variance, and Optimal Risk portfolio for out of sample*

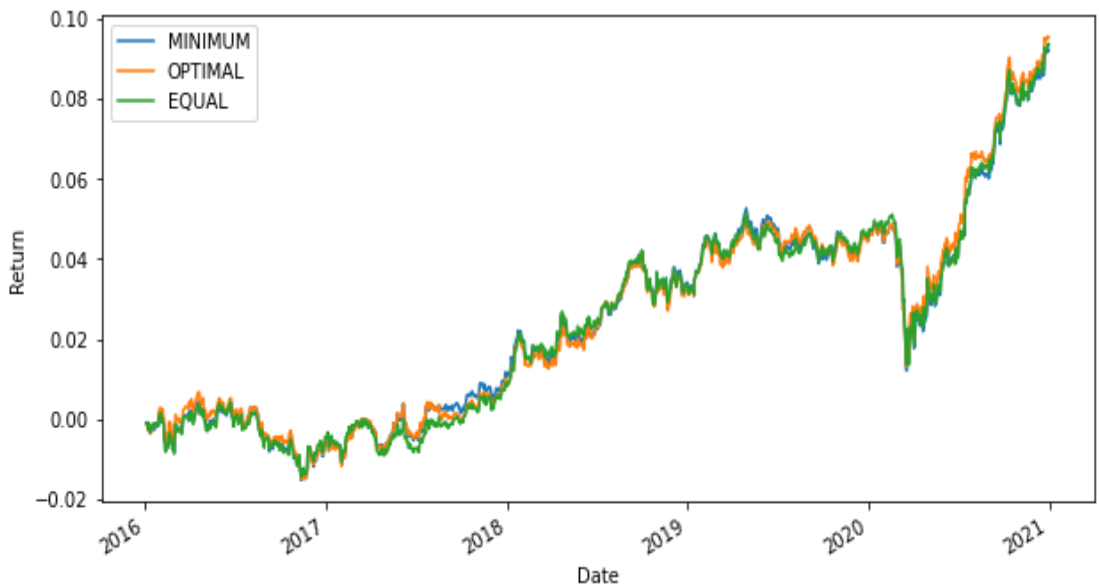

*Figure 0.11: Comparison of Return% of Equal Weight, minimum variance, and Optimal Risk portfolio for out of sample*

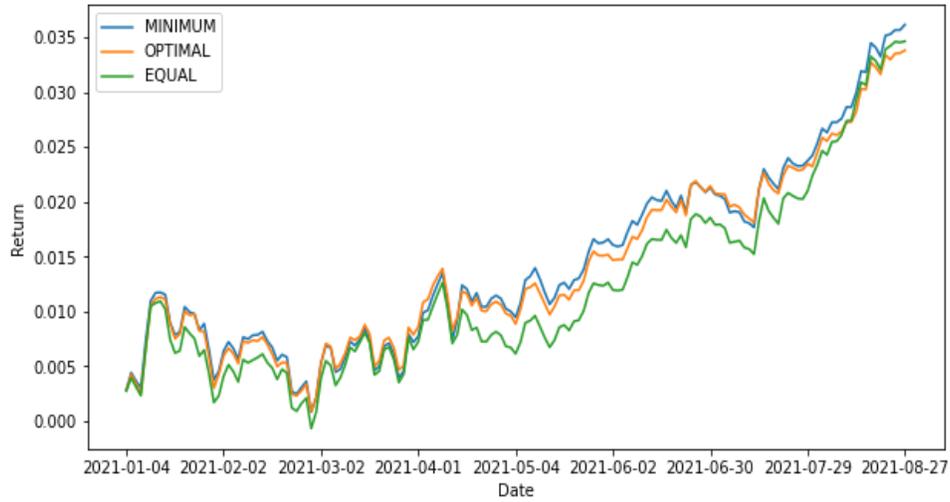

*Table 0.17: In sample results*

| Portfolio | Return | Stdev | Sharpe Ratio |
|-----------|--------|-------|--------------|
| **MINIMUM** | 0.0011 | 0.020 | 0.90 |
| **OPTIMAL** | 0.0012 | 0.021 | 0.91 |
| **EQUAL** | 0.0012 | 0.021 | 0.89 |

*Table 0.18: Out of sample results*

| Portfolio | Return | Stdev | Sharpe Ratio |
|-----------|--------|-------|--------------|
| **MINIMUM** | 0.0035 | 0.0210 | 2.69 |
| **OPTIMAL** | 0.0033 | 0.0200 | 2.64 |
| **EQUAL** | 0.0034 | 0.0206 | 2.631 |

The graph (Fig 6.12) shows that both the line (portfolio return percentage calculated on actual and predicted stock price) closely follow each other which shows the precision in the prediction of stock price

*Figure 0.12: Comparison of Return% of Optimal Risk portfolio on actual stock prices and predicted stock prices for IT sector*

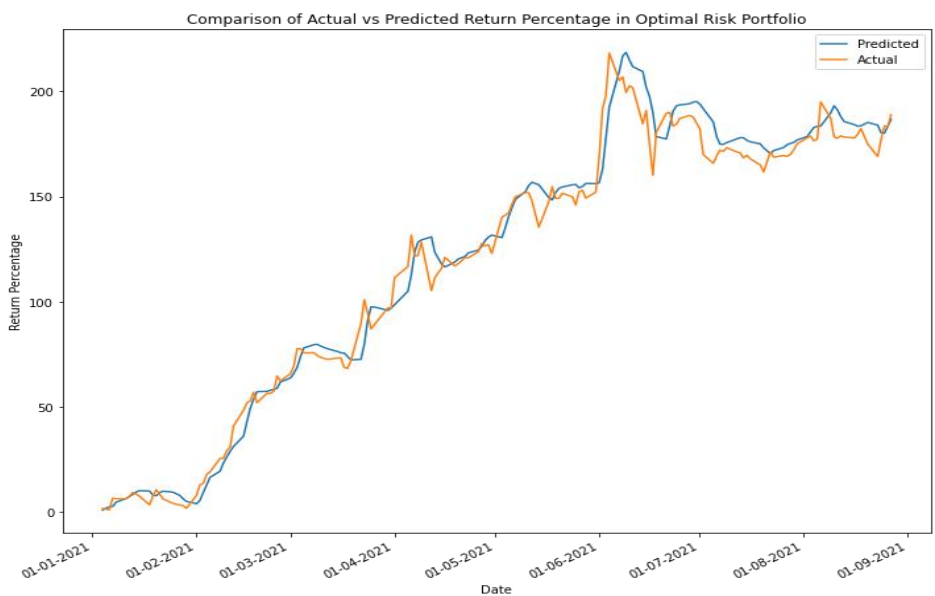

### 4.2.4  Banking Sector

*Table 0.19: Return and risk of banking sector stocks*

| Stocks | Annual Return (%) | Annual Risk (%) |
|---|---|---|
| HDFC Bank | 25.34 | 23.62 |
| ICICI Bank | 24.72 | 36.67 |
| State Bank of India | 3.62 | 37.80 |
| Kotak Mahindra Bank | 29.31 | 28.31 |
| Axis Bank | 9.79 | 38.87 |

Table 6.19 depicts the annual return and the annual risk (i.e. volatility) for the stocks of the banking sector over the training period, i.e., from 1 January 2016 to 27 December 2020. State Bank of India is found to exhibit the lowest annual return and HDFC Bank the lowest annualized risk. Kotak Mahindra Bank yields the highest annual return and Axis Bank the highest risk.

Table 6.20 presents the allocation of weights to different stocks of the banking sector using two portfolio design approaches – (i) minimum risk portfolio and (ii) optimum risk portfolio. The stock which receives the highest allocation of weight as per the minimum risk and optimum risks are HDFC Bank and Kotak Mahindra bank respectively.

*Table 0.20: The portfolios of the banking sector stocks*

| Stocks | Min Risk Portfolio | Opt. Risk Portfolio |
|---|---|---|
| HDFC Bank | 0.55 | 0.38 |
| ICICI Bank | 0.01 | 0.12 |
| State Bank of India | 0.12 | 0.01 |
| Kotak Mahindra Bank | 0.26 | 0.47 |
| Axis Bank | 0.05 | 0.01 |

Table 6.21 shows the risk and the return values associated with the two portfolios of the banking sector stocks. These values are computed using the prices of the stocks over the training period. It is found that between the two portfolios, the optimum risk portfolio yielded the highest values for the return and risk.

*Table 0.21: The return and the risk values of the banking sector portfolios*

| Metric | Min Risk Portfolio | Opt. Risk Portfolio |
|---|---|---|
| Portfolio Return | 22.84% | 26.93% |
| Portfolio Risk | 22.60% | 23.50% |

Table 6.22 shows the return for an investor who followed the optimum risk portfolio approach and invested a total amount of INR 100000 on 1 January 2021. The return for the eight months' period under the optimum risk portfolio is found to be 1.22%. Figure 6.10 exhibits the efficient frontier of the portfolios of the IT sector stocks.

*Table 0.22: The actual return of the optimum portfolio of the banking sector*

| Stock | Date: January 1, 2021 | | | Date: August 31, 2021 | | Return |
|---|---|---|---|---|---|---|
| | Price/Stock | Amount Invested | Volume of Stocks | Price/Stock | Actual Value of Stocks | |
| HDFC Bank | 1425 | 38304.86 | 26.88 | 1548 | 41611.18 | |
| ICICI Bank | 528 | 12787.6 | 24.21 | 700 | 16953.25 | 1.22% |
| State Bank of India | 279 | 883.40 | 3.16 | 412 | 1304.52 | |
| Kotak Mahindra Bank | 1994 | 47811.28 | 23.97 | 1714 | 41097.56 | |
| Axis Bank | 624 | 212.84 | 0.34 | 752 | 256.51 | |

The figure below (Fig 6.13) exhibits the efficient frontier of the portfolios of the banking sector stocks

*Figure 0.13: The minimum risk portfolio (the red star) and the optimum risk portfolio (the green star) for the banking sector on historical stock prices from 1 January 2016 to 27 December 2020 (The risk is plotted along the x-axis and the return along the y-axis)*

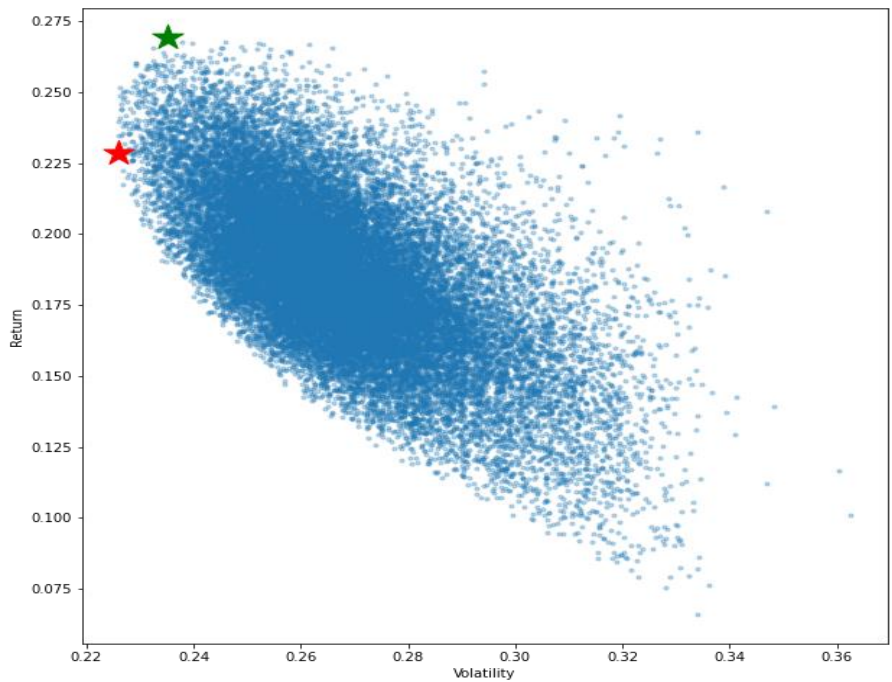

The figure (Fig 6.14) depicts the returns of all the three portfolios during the training period. The graph shows that the optimal risk portfolio obtained returns higher than the minimum variance portfolio and the equal weight portfolio, as expected. The comparison of Sharpe ratio from Table 6.23 validates the same. In the figure (Fig 6.15) backtesting of the portfolios built is depicted. The graph shows that the optimal risk portfolio did not even perform as good as the equal weight portfolio or minimum variance portfolio, which can be validated by comparing the Sharpe ratio of the portfolios from Table 6.24.

*Figure 0.14: Comparison of Return% of Equal Weight, minimum variance, and Optimal Risk portfolio for in sample*

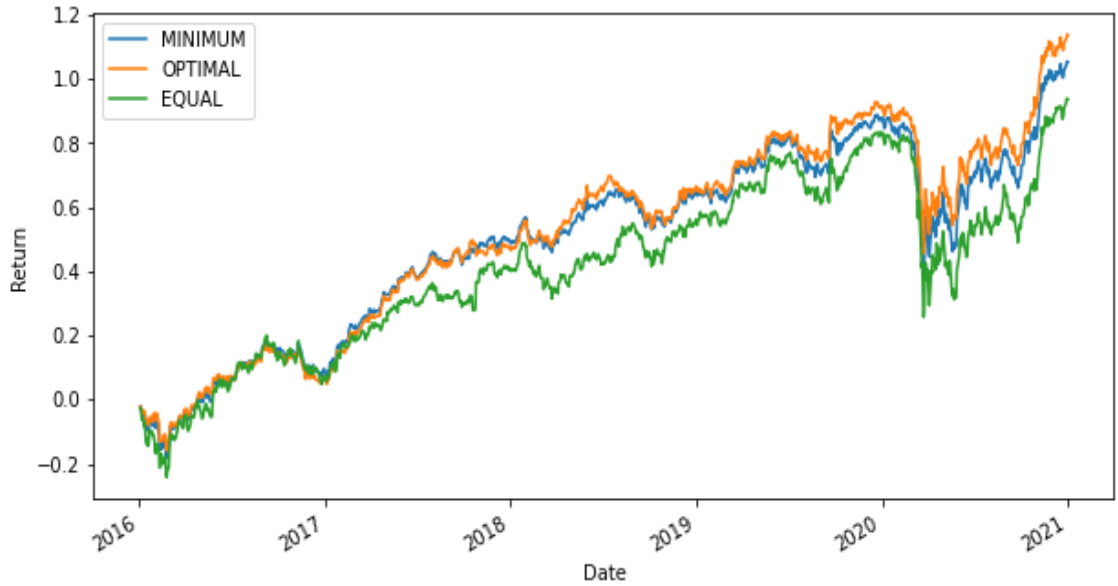

*Figure 0.15: Comparison of Return% of Equal Weight, minimum variance, and Optimal Risk portfolio for out of sample*

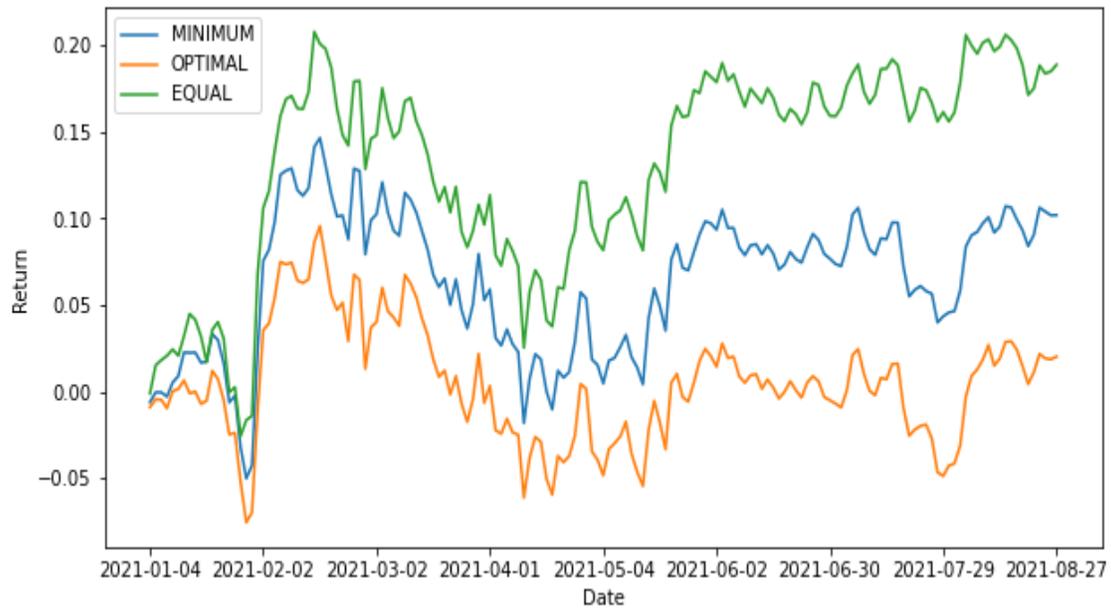

*Table 0.23: In sample results*

| Portfolio | Return | Stdev | Sharpe Ratio |
|-----------|--------|-------|--------------|
| **MINIMUM** | 0.0035 | 0.0210 | 2.69 |
| **OPTIMAL** | 0.0033 | 0.0200 | 2.64 |
| **EQUAL** | 0.0034 | 0.0206 | 2.63 |

*Table 0.24: Out sample results*

| Portfolio | Return | Stdev | Sharpe Ratio |
|-----------|--------|-------|--------------|
| **MINIMUM** | 10.05 | 0.24 | 0.65 |
| **OPTIMAL** | 1.99 | 0.23 | 0.13 |
| **EQUAL** | 18.61 | 0.25 | 1.15 |

The graph (Fig 6.16) shows that both the line (portfolio return percentage calculated on actual and predicted stock price) closely follow each other which shows the precision in the prediction of stock price

*Figure 0.16: Comparison of Return% of Optimal Risk portfolio on actual stock prices and predicted stock prices for banking sector*

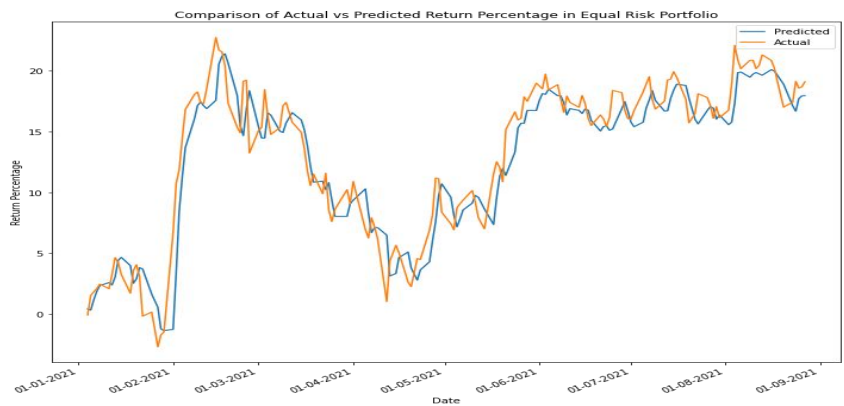

### 4.2.5 Auto sector

Table 6.25 depicts the annual return and the annual risk (i.e. volatility) for the stocks of the auto sector over the training period, i.e., from 1 January 2016 to 27 December 2020. Tata Motors is found to exhibit the lowest annual return and Bajaj the lowest annualized risk. Maruti Suzuki yields the highest annual return and Tata Motors the highest risk.

*Table 0.25: Return and risk of auto sector stocks*

| Stocks | Annual Return (%) | Annual Risk (%) |
|---|---|---|
| Maruti Suzuki | 15.53686 | 31.37358 |
| Tata Motors | -15.5022 | 47.35468 |
| Mahindra & Mahindra | 8.878597 | 31.80773 |
| Bajaj | 8.366865 | 26.10789 |
| Eicher Motors | 6.279331 | 34.32935 |

*Table 0.26: The portfolios of the auto sector stocks*

| Stocks | Min Risk Portfolio | Opt. Risk Portfolio |
|---|---|---|
| Maruti Suzuki | 0.11 | 0.79 |
| Tata Motors | 0.02 | 0.01 |
| Mahindra & Mahindra | 0.23 | 0.15 |
| Bajaj | 0.51 | 0.01 |
| Eicher Motors | 0.11 | 0.04 |

Table 6.26 presents the allocation of weights to different stocks of the banking sector using two portfolio design approaches – (i) minimum risk portfolio and (ii) optimum risk portfolio. The stock which receives the highest allocation of weight as per the minimum risk and optimum risks are Bajaj and Maruti Suzuki respectively.

Table 6.27 shows the risk and the return values associated with the two portfolios of the auto sector stocks. These values are computed using the prices of the stocks over the training period. It is found that between the two portfolios, the optimum risk portfolio yielded the highest values for the return and risk.

*Table 0.27: The return and the risk values of the auto sector portfolios*

| Metric | Min Risk Portfolio | Opt. Risk Portfolio |
|---|---|---|
| Portfolio Return | 8.50% | 14.00% |
| Portfolio Risk | 23.24% | 28.81% |

Table 6.28 shows the return for an investor who followed the optimum risk portfolio approach and invested a total amount of INR 100000 on 1 January 2021. The return for the eight months' period under the optimum risk portfolio is found to be -9.94%.

*Table 0.28: The actual return of the optimum portfolio of the auto sector*

| Stock | Date: January 1, 2021 | | | Date: August 31, 2021 | | Return |
|---|---|---|---|---|---|---|
| | Price/ Stock | Amount Invested | Volume of Stocks | Price/Stock | Actual Value of Stocks | |
| Maruti Suzuki | 7691 | 79715.32 | 10.36 | 6625 | 68666.49 | |
| Tata Motors | 186 | 253.67 | 1.36 | 286 | 390.06 | -9.94% |
| Mahindra & Mahindra | 732 | 15283.09 | 20.87 | 775 | 16180.86 | |
| Bajaj | 3481 | 177.80 | 0.05 | 3699 | 188.94 | |
| Eicher Motors | 2543 | 4570.11 | 1.79 | 2576 | 4629.41 | |

The figure below (Fig 6.17) exhibits the efficient frontier of the portfolios of the auto sector stocks.

*Figure 0.17: The minimum risk portfolio (the red star) and the optimum risk portfolio (the green star) for the auto sector on historical stock prices from 1 January 2016 to 27 December 2020 (The risk is plotted along the x-axis and the return along the y-axis)*

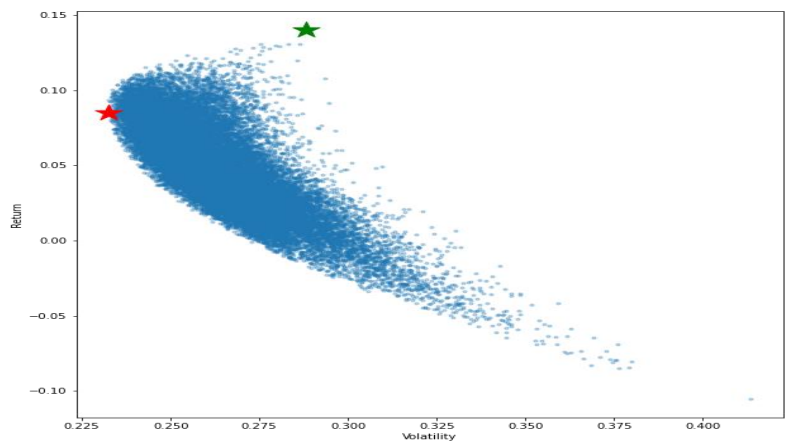

The figure (Fig 6.18) depicts the returns of all the three portfolios during the training period. The graph shows that the optimal risk portfolio obtained returns higher than the minimum variance portfolio and the equal weight portfolio, as expected. The performance of the portfolios can be better compared using Sharpe ratio of the portfolios from Table 6.29. In the figure (Fig 6.19) backtesting of the portfolios built is depicted. The graph shows that the optimal risk portfolio was not as good as the equal weight portfolio or minimum variance portfolio, which can be validated by comparing the Sharpe ratio of the portfolios from Table 6.30.

*Figure 0.18: Comparison of Return% of Equal Weight, minimum variance, and Optimal Risk portfolio for in sample*

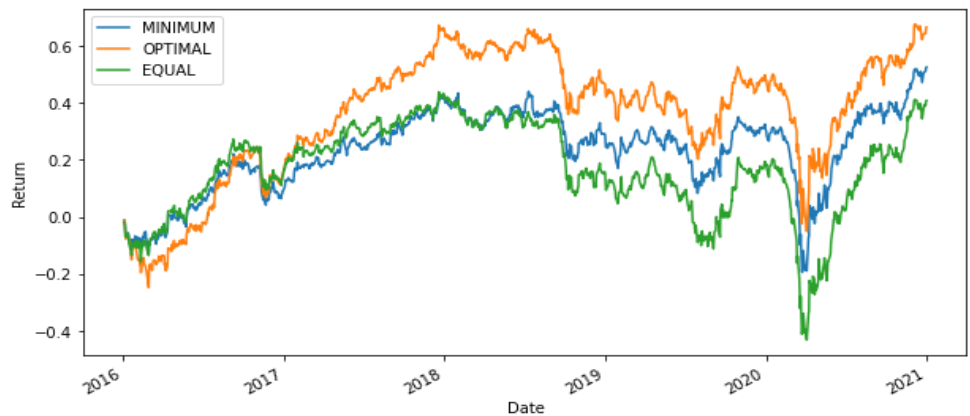

*Figure 0.19: Comparison of Return% of Equal Weight, minimum variance, and Optimal Risk portfolio for out of sample*

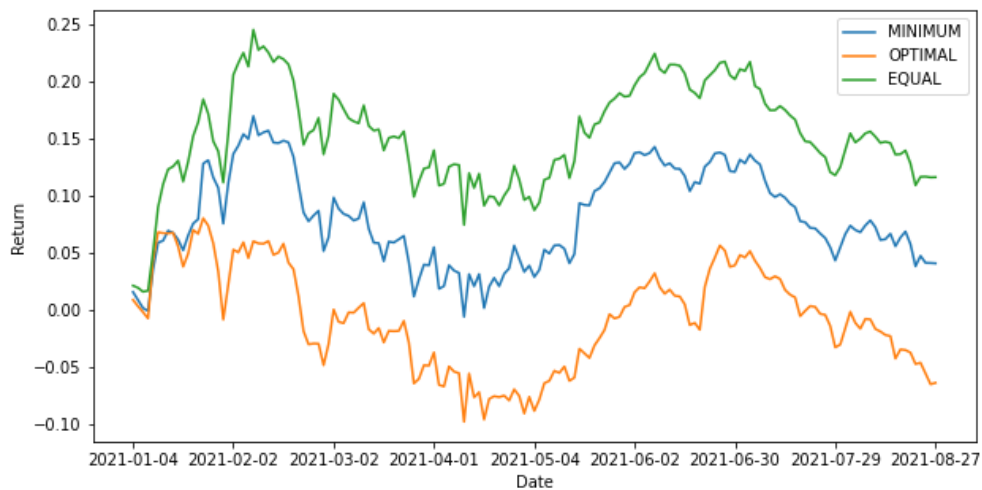

*Table 0.29: In sample results*

| Portfolio | Return | Stdev | Sharpe Ratio |
|-----------|--------|-------|--------------|
| **MINIMUM** | 6.78 | 0.23 | 0.46 |
| **OPTIMAL** | 8.58 | 0.26 | 0.50 |
| **EQUAL** | 5.26 | 0.25 | 0.32 |

*Table 0.30: Out of sample results*

| Portfolio | Return | Stdev | Sharpe Ratio |
|-----------|--------|-------|--------------|
| **MINIMUM** | 4.02 | 0.21 | 0.29 |
| **OPTIMAL** | -6.29 | 0.21 | -0.45 |
| **EQUAL** | 11.45 | 0.24 | 0.75 |

The graph (Fig 6.20) shows that both the lines (portfolio return percentage calculated on actual and predicted stock price) closely follow each other which shows the precision in the prediction of stock price.

*Figure 0.20: Comparison of Return% of Optimal Risk portfolio on actual stock prices and predicted stock prices for auto sector*

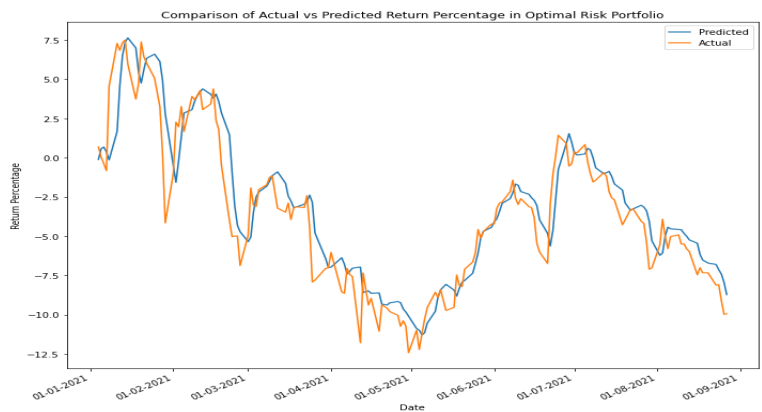

The results show that only the metal sector gave expected results during the back testing. In the case of the IT sector, during the training period itself Optimal portfolio failed to give good results due to the high correlation between the stocks involved. For all the other three sectors, during the testing phase optimal risk portfolio has not performed as good as minimum variance portfolio and in some cases not even as good as equal weight portfolio. This point towards the problem of overfitting the train data in the optimal risk portfolio technique. Since the weights are calculated based on the training data and as stock prices are highly dynamic, overfitting the train data can give erroneous results in the testing phase. Employing regularization techniques is the way out and in the case of stocks it is diversification of stocks in the portfolio. Advanced portfolio optimization techniques should be employed in this regard.

# Chapter 7

## Conclusion

In this work, we have studied the performance of statistical, econometrical, machine learning and deep learning models in stock price prediction. Linear regression is the statistical model studied. ARIMA, MARS and VAR are the econometric models studied whereas K Nearest Neighbor, Decision Tree, SVM, XGBoost and Random Forest are the machine learning models used for stock price prediction. Among the classification ML models, along with the above mentioned ones, logistic regression has also been studied. LSTM and CNN are the deep learning models used for stock price prediction. Performance of the deep learning models with increase or decrease of the parameters/ input values has also been analyzed. Walk Forward Validation is the validation method used for statistical, econometric and machine learning models. Since deep learning models give better results only if the training data is huge, traditional train slit method is used for training and testing. Some interesting patterns in day wise RMSE/ mean plots have been noticed while building machine learning and deep learning models. It is noticed that among the econometric models MARS performed the best across the sectors, whereas Random Forest and XGBoost are at par with each other as far as ML classification or regression models are concerned. LSTM gave better results compared to CNN, though time taken for execution is much less for CNN.

Minimum variance portfolio and optimal risk portfolio are the two portfolio building methods used. Training period employed is 5 years for building the portfolio and the back testing is done for the next 8 months. An analysis of the performance of both the portfolios is done by comparing it with the equal weighted portfolio based on the back testing results. Interestingly, it was noticed that except for the metal sector for all other 4 sectors the performance of the optimal weight portfolio is not even at par with the minimum variance portfolio. This shows the shortcomings of the optimal risk portfolio and points towards the need of better portfolio optimization techniques that would not overfit the training data. Various advanced optimization techniques like HERC (Hierarchical Equal Risk Contribution), HRP (Hierarchical Risk Parity) etc. ensures regularization by diversification of the portfolio. We believe that travelling further in the road of portfolio optimization techniques would bring in new insights and we look forward to the same.